\documentclass[aps,prd,eqsecnum,11pt,preprintnumbers,nofootinbib]{revtex4-1}
\pdfoutput=1

\def\nbox#1#2{\vcenter{\hrule \hbox{\vrule height#2in
\kern#1in \vrule} \hrule}}
\def\sq{\,\raise.5pt\hbox{$\nbox{.09}{.09}$}\,}
\def\sqb{\,\raise.5pt\hbox{$\overline{\nbox{.09}{.09}}$}\,}

\def\XXint#1#2#3{{\setbox0=\hbox{$#1{#2#3}{\int}$}
\vcenter{\hbox{$#2#3$}}\kern-.5\wd0}}

\newcommand{\bea}{\begin{eqnarray}}
\newcommand{\eea}{\end{eqnarray}}
\newcommand{\be}{\begin{equation}}
\newcommand{\ee}{\end{equation}}
\newcommand{\bes}{\begin{subequations}}
\newcommand{\ees}{\end{subequations}}

\def\eps{\epsilon}

\def\nn{\nonumber\\}

\def\nbox#1#2{\vcenter{\hrule \hbox{\vrule height#2in
\kern#1in \vrule} \hrule}}
\def\sq{\,\raise.5pt\hbox{$\nbox{.09}{.09}$}\,}
\def\sqb{\,\raise.5pt\hbox{$\overline{\nbox{.09}{.09}}$}\,}

\usepackage{setspace}
\usepackage{amssymb,amsmath}
\usepackage{graphicx}
\usepackage{graphics}
\usepackage{sidecap}
\usepackage{enumitem}

\allowdisplaybreaks

\begin{document}

\preprint{LA-UR-15-20030}

\title{Surface Tension  and Negative Pressure Interior of a Non-Singular `Black Hole'}
\author{Pawel O. Mazur}
\affiliation{Department of Physics and Astronomy\\
University of South Carolina\\
Columbia, SC 29208 USA}
\email{mazur@caprine.physics.sc.edu}
\author{Emil Mottola}
\affiliation{Theoretical Division, T-2 \\
Los Alamos National Laboratory \\
Los Alamos, NM 87545 USA}
\email{emil@lanl.gov}

\begin{abstract}
\vspace{2mm}

The constant density interior Schwarzschild solution for a static, spherically symmetric collapsed star has a 
divergent pressure when its radius $R \le \frac{9}{8} R_s =\frac{9}{4}GM$. We show that this divergence 
is integrable,  and induces a non-isotropic transverse stress with a finite redshifted surface tension 
on a spherical surface of radius $R_0 = 3R \sqrt{1 - \frac{8}{9}\!\frac{R\,}{R_s}}$. For $r < R_0$ the 
interior Schwarzschild solution exhibits negative pressure. When $R = R_s$, the surface is localized at 
the Schwarzschild radius itself, $R_0=R_s$, and the solution has constant negative pressure $p = - \bar\rho$ 
everywhere in the interior $r< R_s$, thereby describing a gravitational condensate star, a fully collapsed 
non-singular state already inherent in and predicted by classical General Relativity. The redshifted surface 
tension of the condensate star surface is given by $\tau_s = \Delta \kappa/8 \pi G$, where 
$\Delta \kappa = \kappa_+ - \kappa_- = 2 \kappa_+ = 1/R_s$ is the difference of equal and opposite surface 
gravities between the exterior and interior Schwarzschild solutions. The First Law,  
$dM = dE_v + \tau_s\, dA$ is recognized as a purely mechanical classical relation at zero temperature 
and zero entropy, describing the volume energy and surface energy change respectively. Since there is no 
event horizon, the Schwarzschild time $t$ of such a non-singular gravitational condensate star is a global time, 
fully consistent with unitary time evolution in quantum theory. The $p=-\bar\rho$ interior acts as 
a defocusing lens for light passing through the condensate, leading to imaging 
characteristics distinguishable from a classical black hole. A further observational test of gravitational condensate 
stars with a physical surface vs. black holes is the {\it discrete} surface modes of oscillation which
should be detectable by their gravitational wave signatures.
\end{abstract}
\maketitle

\section{Introduction}
\label{Sec:Intro}

The endpoint of complete gravitational collapse of a star whose mass exceeds some critical value of order
of a few solar masses is widely assumed to be a singular state called a `black hole.' This hypothesis rests
upon the fact that continued gravitational collapse past nuclear densities cannot be halted by any conventional
equation of state for high density matter. If a trapped surface from which light cannot escape is formed,
and the energy density of the matter $\rho$ plus the sum of the principal pressures $p_i$ is positive
\vspace{-5mm}
\be
\rho + \sum_{i=1}^3 p_i  \ge 0
\label{strongcond} 
\ee
then the Einstein equations of classical General Relativity are sufficient to prove that gravitational collapse must 
result in a singularity, and black hole formation is inevitable  \cite{HawPen,HawEll}.

The strong energy condition (\ref{strongcond}) holds for all known forms of matter or radiation, but for one 
significant exception, {\it viz.} the cosmological vacuum dark energy believed to be responsible for the 
accelerating expansion of the universe. The equation of state of vacuum dark energy is $p = -\bar\rho <0$ 
with constant {\it negative} pressure, so that $\rho + 3p = -2\bar\rho < 0$. Indeed, it is just this inequality that 
results in the defocusing rather than the focusing of geodesics, which is the effective repulsive force presumed 
to be causing the expansion of the universe  to accelerate \cite{SNI}. If such an equation of state, violating 
(\ref{strongcond}) were to be realized within the central regions of a star undergoing gravitational collapse, 
then the same effective repulsive force could prevent the formation of a black hole singularity.

Had cosmological dark energy and the accelerating expansion of the universe not been discovered 
observationally, other considerations lead naturally to this same $p= -\bar\rho <0 $ equation of state.
In classical General Relativity it is the eq. of state of a (positive) cosmological term. In quantum theory 
vacuum fluctuations lead generically to a non-zero cosmological term, resulting in this same eq. of state \cite{AMM07}.
The vacuum energy density with $p= - \bar\rho$  due to gluon condensation appears in the bag model for hadrons 
and is necessary to obtain agreement both with QCD sum rules and with experiment \cite{ShiVaiZak}. 
Other violations of (\ref{strongcond}) are predicted by quantum theory dependent upon external conditions
such as in the Casimir effect, now confirmed by a number of laboratory experiments \cite{BorMohMil}. It is only 
the extreme feebleness of of the gravitational interaction that prevents the gravitational effects of these 
expected violations of the strong energy condition from being directly observed in laboratory conditions.

Perhaps most persuasive of all, quantum theory applied to black hole spacetimes, as they are usually described, 
possessing an event horizon and an interior singularity, leads to a number of severe paradoxes associated 
with the validity of unitary time evolution and the preservation of information apparently `lost' by quantum 
matter falling into a black hole \cite{Hawunit}. Certainly if macroscopic matter disappears into a singularity 
of space and time all predictive power of present physical theories, whether quantum or classical is lost as well. 
Causality and the classical singularity theorems assure us that this common fate of both matter and otherwise 
successful theories such as the Standard Model is inevitable once the event horizon is crossed, provided 
(\ref{strongcond}) holds everywhere within it. Avoiding this fate and removing these difficulties consistently
within quantum theory and General Relativity requires the abandonment of (\ref{strongcond}), and as 
a result, some modification of both black hole interiors and/or horizons. 

The singularity theorems illustrate the power of arguments and the broad conclusions that may be drawn 
in Einstein's theory based on simple general bounds or inequalities obeyed by the matter stress-energy, 
such as (\ref{strongcond}), otherwise independently of its detailed composition or eq. of state. When 
spherical symmetry is assumed, it is possible to prove an even more remarkable set of theorems about 
the existence or non-existence of stable fully collapsed stars, relying on apparently even weaker conditions 
on the matter stress-energy. Assuming only isotropic pressures ($p_i = p$ for all $i=1, 2, 3 $), a positive 
energy density that is a non-increasing function of radius,  $\displaystyle \frac{d\rho}{dr} \le 0$\,, and matching 
the interior of a star to the vacuum Schwarzschild solution exterior to it, it was shown by Buchdahl \cite{Buch}
that the radius of the star must be greater than a certain finite value, $ R > \frac{9}{8} R_s$, with $R_s = 2 GM$ 
the Schwarzschild radius, in order for the interior solution to possess everywhere finite pressure: {\it cf.} Appendix \ref{Sec:Buchdahl}. 
Subsequently, it was shown that a rigorous lower bound on the central pressure can be established under the same 
assumptions \cite{MarVis}. These theorems do not require the strong energy condition (\ref{strongcond}). The 
existence of these bounds implies that something drastic must happen to the interior of a collapsed star even 
when measurably ({\it i.e.}\,macroscopically) outside of its Schwarzschild radius, {\it before} a trapped surface 
forms, at least under the assumptions of spherical symmetry, continuity and monotonicity $\rho' \le 0$, entirely 
within the domain of classical General Relativity.

Moreover, the Buchdahl and related bounds are established and saturated by comparison to the Schwarzschild 
interior solution with $\rho' = 0$, {\it i.e.}\,$\rho = \bar\rho =$ constant \cite{Schw2,Nar}. This constant density interior 
Schwarzschild solution has a divergent central pressure when its radius first reaches the Buchdahl bound 
$R = \frac{9}{8} R_s$. The assumption of strictly constant density, presumed unphysical, together with this pressure 
divergence of the interior Schwarzschild solution has usually been regarded as reasons enough to exclude it from 
further consideration \cite{Wein,Wald}. As a result, the behavior of the solution for  $R \le \frac{9}{8} R_s$ has been 
little studied or remarked upon in the literature \cite{CatFabVis}. However, the existence of the Buchdahl and related 
bounds in which the constant density solution is the limiting case, makes the study of the interior Schwarzschild solution 
relevant to and potentially quite instructive for the general case and realistic application to any spherically symmetric 
fully collapsed self-gravitating mass.

When one does consider the constant density interior Schwarzschild solution for $R < \frac{9}{8} R_s$,
some of its rather remarkable features quickly become apparent. First, the pressure divergence moves out from 
the origin to a spherical surface of finite radius 
\be
R_0 = 3R\, \sqrt{1 - \frac{8}{9}\frac{R\,}{R_s}} >0   \qquad {\rm for} \qquad R < \frac{9}{8} R_s = \frac{9}{4}\, GM
\label{R0}
\ee
and a {\it new regular solution} for $0\le r < R_0$ opens up behind it, with {\it negative} pressure, violating (\ref{strongcond}). 
Second, as the radius of the star approaches the Schwarzschild radius from above $R \rightarrow R_s^+$, this region of 
negative pressure moves outward from the origin and comes to encompass the entire interior, since $R_0 \rightarrow R_s^-$ 
from below. Third, most remarkably of all, in the above limit the entire interior becomes one of {\it constant 
negative pressure} with precisely the $p = - \bar \rho$ de Sitter dark energy equation of state. Because of the Buchdahl 
bound and related theorems one can conclude that negative pressure and the dark energy `quantum vacuum' 
equation of state are already inherent in and predicted to occur in classical General Relativity quite generally,
under sufficiently severe conditions of spherically symmetric gravitational compression, and prior to any formation
of a trapped surface or event horizon.

The crucial feature of the pressure singularity at $r=R_0$ of the interior Schwarzschild solution for $R < \frac{9}{8} R_s$ 
is that the norm of the static Killing vector vanishes at the {\it same} point, touching zero in a cusp-like behavior at $r=R_0$, 
but otherwise remaining positive ({\it cf.} Figs.\,\ref{Fig:Redshiftsing}-\ref{Fig:hvarious} and Figs.\,\ref{Fig:Redshiftfinal}-\ref{Fig:hfinal}).
In this paper we show that the pressure singularity and cusp is in fact integrable through the Komar 
integral formula for the total mass-energy of a stationary configuration \cite{Tolm,Komar}, and results in a distributional 
$\delta$-function in the transverse stress $T^{\theta}_{\ \theta} =T^{\phi}_{\ \phi} \equiv p_{\perp} \neq T^r_{\ r}$ 
localized precisely at $r=R_0$. Thus in the limit $R \rightarrow R_s^+, R_0 \rightarrow R_s^-$ the classical interior 
Schwarzschild solution describes a non-singular gravitational condensate star with a physical surface and finite 
surface tension, proposed in \cite{gstar,PNAS}  (see also \cite{CHLS}), in which the thickness of the thin shell 
quantum surface layer goes to zero. 

The surface tension may be computed in terms of the discontinuity of equal magnitude and opposite 
signed surface gravities $\kappa_{\pm} = - \kappa_{\mp}$ at $r=R_0$, and in fact, this example serves
to generalize the Israel junction conditions \cite{Isr,PoiIsr} to a null boundary layer in an unambiguous way. 
In terms of the (redshifted) surface tension and corresponding surface energy to be determined in Sec. \ref{Sec:Tension}
the differential First Law of Black Hole Mechanics \cite{Smar,BarCarHaw} is straightforwardly modified and 
recognized to be a purely mechanical classical relation between volume and surface contributions to the 
Komar energy. There is no temperature or entropy whatsoever associated with the limiting gravitational 
condensate star configuration, as both are identically zero. Moreover, since the Schwarzschild Killing time 
extends throughout the interior and exterior Schwarzschild solution including for $R_0=R=R_s$, and as 
there is neither a spacetime singularity nor an event horizon, quantum mechanical evolution of fields in 
such a background static spacetime is clearly unitary.

The paper is organized as follows. In the next section we review the constant density interior 
Schwarzschild solution and show how negative pressures appear when $R < \frac{9}{8} R_s$,
summarizing the main results of the paper. The succeeding sections starting with Sec. \ref{Sec:Mass}
assemble the necessary formalism to analyze the general stationary configuration of matter and gravity 
through the Komar mass-energy integral, energy flux and surface gravity. In Sec. \ref{Sec:Tension} we
apply the Komar formula to the interior Schwarzschild solution and show that its pressure singularity 
is integrable, and corresponds precisely to a positive transverse pressure $p_{\perp} - p > 0$ and 
surface tension $\tau_s$ localized on the discontinuous pressure surface. In Sec. \ref{Sec:FirstLaw} 
we show how the First Law of energy conservation applied to the Schwarzschild interior solution 
may be recognized as a purely mechanical classical relation of volume and areal surface energy, 
the latter determined by the surface tension, at strictly zero temperature and entropy. 
In Sec. \ref{Sec:Observ} two properties of negative pressure gravitational condensate stars with
a physical surface that may permit them to be distinguished observationally from black holes
are discussed, {\it viz.} the defocusing of null rays passing through the interior leading to different
optical imaging properties, and the excitation of discrete frequency surface modes detectable by their 
gravitational wave signatures. Sec. \ref{Sec:Conclusions} contains our Conclusions and a Discussion 
of extension of the simple model presented to more general situations and its embedding in a more
complete theory.

There are three Appendices. In Appendix \ref{Sec:Buchdahl} the Buchdahl and related bounds are
summarized. Appendix \ref{Sec:DeltaDistrib} contains the mathematical details of how the $\delta$-function 
distribution in the transverse pressure may be obtained by a careful regulation of the pressure singularity
by a small parameter $\eps$ in the limit $\eps \rightarrow 0^+$, suggesting also how a physical regulator 
$\eps \propto \sqrt{\hbar}$ dependent upon quantum corrections may enter in a more complete quantum theory. 
Finally in Appendix \ref{Sec:Junction} the relationship between the Israel junction conditions 
for a spacelike boundary surface as assumed in \cite{PNAS} and the limit in which the boundary 
between the modified de Sitter interior and Schwarzschild exterior becomes null is explained.

\section{Interior Schwarzschild Solution and Negative Pressure}
\label{Sec:Interior}

The general static, spherically symmetric line element in Schwarzschild coordinates is
\be
ds^2 = -f(r)\, dt^2 + \frac{dr^2}{h(r)} + r^2\left( d\theta^2 + \sin^2\theta\,d\phi^2\right)
\label{sphsta}
\ee
in terms of two metric functions $f(r)$ and $h(r)$. The Schwarzschild time coordinate $t$ is invariantly
defined by the existence of a Killing vector $K^{\mu}$ such that
\be
\frac{\partial}{\partial t} = K^{\mu}\frac{\partial}{\partial x^{\mu}}\,,\qquad\ K^{\mu} = \delta^{\mu}_{\ t}
\label{Killt}
\ee
and the geometry is independent of $t$. From (\ref{sphsta}) and (\ref{Killt}) we see that
\be
-K^{\mu}K_{\mu} = -g_{tt} = f(r)
\label{Killnorm}
\ee
is a scalar invariant. The radius $r$ is similarly defined in an invariant geometric manner 
by the condition that $A= 4 \pi r^2$ is the area of the spherical two-surface at fixed $r$ and fixed $t$.

The form of the stress-energy tensor of a general static, spherically symmetric 
distribution of matter may be expressed as the diagonal matrix
\vspace{-1mm}
\be
T^{\mu}_{\ \,\nu} = \left(\begin{array}{cccl} -\rho & 0 & 0 & 0\\
0 & p & 0 & 0\\ 0 & 0 & p_{\perp} & 0\\ 0 & 0 & 0 & p_{\perp} \end{array}\right) 
\label{Tgeneral}
\ee
in the Schwarzschild coordinates $(t, r, \theta, \phi)$. The three functions $\rho, p$ and $p_{\perp}$ are the 
mass-energy density, radial pressure, and tangential pressure respectively. Thus the general, spherically
symmetric static configuration of matter and geometry requires five functions of $r$ in all. 

These five functions are required to satisfy the Einstein eqs. of classical General Relativity, whose independent 
information is contained in the two components,
\bes
\begin{align}
-G_{\ t}^t &=  \frac{1}{r^2} \frac{d}{dr}\left[r\left(1 - h\right)\right]
= -8\pi G\, T_{\ t}^t = 8\pi G \rho \label{Gtt}\\
G_{\ r}^r &= \frac{h}{r f}\frac{d f}{dr}  + \frac{1}{r^2} \left(h -1\right) =
8\pi G \,T_{\ r}^r = 8\pi G p 
\label{Einp}
\end{align}
\label{Eins}\ees
together with the covariant conservation eq.
\be
\nabla_{\mu}\, T^{\mu}_{\ r} = \frac{d p}{dr} + \frac{\rho + p}{2f} \,\frac{d f}{dr}  + \frac{2\, (p - p_{\perp})}{r}= 0
\label{cons}
\ee
which expresses the pressure balance of forces in static equilibrium.
It is sometimes convenient to trade the two metric functions $f(r)$ and $h(r)$ for the 
gravitational potential $\Phi(r)$ and the Schwarzschild or Misner-Sharp mass  \cite{MisSharp} 
within a sphere of radius $r$, $m(r)$, defined by 
\be
f = e^{2 \Phi}\,,\qquad h = 1 - \frac{2Gm}{r}
\label{Phimdef}
\ee
respectively. From (\ref{Gtt}) $\displaystyle \frac{dm}{dr} = 4 \pi r^2 \rho$
so that $m(r)$ is obtained from this by direct integration 
\be
m(r) = 4 \pi\! \int_0^r r^2 \rho(r)\, dr
\label{mint}
\ee
assuming $m(0) =0$. The remaining Einstein eq. (\ref{Einp}) and (\ref{cons}) thereby become
\bes
\bea
&&h\, \frac{d\Phi}{dr} = \frac{Gm}{\,r^2} + 4 \pi G p r  \label{dPhidr}\\
&&\frac{d p}{dr} + (\rho + p) \,\frac{d \Phi}{dr}  = \frac{2 (p_{\perp}- p)}{r} 
\label{pperp}
\eea
\label{dPhi}\ees
so that $\Phi (r)$ is the Newtonian gravitational potential in the non-relativistic limit 
where $p, p_{\perp} \ll \rho$ and $G m/r^2 \ll 1$, $h \approx 1$ (in units in which $c=1$).

\vspace{-1mm}
For a star of total mass $M$ and radius $R$, the metric functions $f$ and $h$ must match the exterior 
Schwarzschild solution {\it in vacuo} 
\be
f_{ext}(r) = h_{ext}(r) = 1 - \frac{2GM}{r} = 1 - \frac{R_s}{r}\,,\qquad R_s \equiv 2GM\,,\qquad r \ge R
\label{extSch}
\ee
where a possible multiplicative constant of integration in $f(r)$ is fixed by the condition that the line 
element (\ref{sphsta}) approach that of flat space Minkowski with the standard interval of time as $r \rightarrow \infty$.
In addition, for equilibrium the pressure must vanish at $r=R$. Thus we have the boundary conditions
\be
p(R) = 0 \,,\qquad m(R) = M\,, \qquad f(R) = h(R) = 1 - \frac{R_s}{R}  
\label{condR}
\ee
for the interior solution at the surface of the star.

With these conditions the interior solution for $r < R$ is still underdetermined, and additional information about
the matter stress tensor must be supplied. Most commonly, perfect isotropic fluid behavior is assumed by setting
$p_{\perp} = p$.  Schwarzschild in his second paper \cite{Schw2} assumed in addition to this perfect fluid behavior 
that the interior has constant density
\be
\rho = \bar \rho \equiv \frac{3M}{4 \pi R^3}
\label{constden}
\ee
and these two additional conditions allow eqs. (\ref{Eins})-(\ref{cons}) to be solved in closed form. 
In that case eq. (\ref{mint}) is integrated immediately to obtain
\be
m(r) = \frac{4 \pi}{3} \bar \rho r^3 = \frac{M} {R^3}\, r^3\,,\qquad h(r) = 1 - H^2 r^2\,,\qquad 0\le r \le R 
\label{mhr}
\ee
where we have defined
\be
H^2 \equiv \frac{8 \pi G}{3} \bar\rho = \frac{2GM}{R^3} = \frac{R_s}{R^3}\,.
\label{defH2}
\ee
Then eliminating the gravitational potential function $\displaystyle\frac{d\Phi}{dr}$ from eqs. (\ref{dPhi}) gives
\be
h\,\frac{d p}{dr} +\frac{ (\bar\rho + p)}{2} \,\left( H^2 r + 8 \pi G p r\right) =  \frac{2 (p_{\perp} - p)}{r} = 0
\label{dpdr}
\ee
where the last equality is valid if $p_{\perp} = p$. In view of (\ref{mhr}),  (\ref{dpdr}) 
may be written in the separable form
\be
\frac{dp}{(p + \bar\rho)(8 \pi G \,p + H^2)} = - \frac{r\ dr}{2\, (1-H^2r^2)}
\ee
whose solution is elementary. Integrating from the outer boundary at $r=R$ where $p(R) =0$ to $r$ gives
\be
\frac{p(r) + \bar \rho}{3p(r) + \bar \rho} = \frac{\sqrt{1- H^2R^2}}{\sqrt{1-H^2r^2}}
\ee
or solving for the pressure,
\be
p(r) = \bar\rho \left[ \frac{ \sqrt{1 - H^2 r^2} - \sqrt{1 - H^2 R^2}} {3 \sqrt{1 - H^2 R^2} - \sqrt{1 - H^2 r^2}} \right]
\label{psoln}
\ee
and we also have
\vspace{-3mm}
\be
p + \bar\rho = 2 \, \bar\rho\, \left[ \frac{\sqrt{1 - H^2 R^2}} {3 \sqrt{1 - H^2 R^2} - \sqrt{1 - H^2 r^2}}\right]
\label{pplusrho}
\ee
for $r \le R$. Lastly the solution for $f$ or $\Phi$ obeying the boundary condition (\ref{condR}) is 
easily found to be
\be
f(r) = e^{2 \Phi} = \frac{1}{4} \left[3  \sqrt{1 - H^2 R^2} - \sqrt{1 - H^2 r^2}\,\right]^2 \ge 0
\label{fsoln}
\ee
completing the constant density Schwarzschild interior solution matched to the vacuum Schwarzschild
exterior solution at $r=R$.

Several remarks about this solution bear emphasis. First, its importance is not due to any assumption
(at this point at least) that constant density $\rho =\bar\rho$ represents a realistic eq. of state for high 
density matter. Rather it represents an extreme situation which can be used as a bound and 
an instructive model for the general spherically symmetric interior solution. Second, since the eqs. 
for $p(r)$ and $f(r)$ are first order, with boundary conditions (\ref{condR}), there is no freedom to 
adjust the first derivative $f' (R)$. However this first derivative is also continuous with the exterior 
Schwarzschild solution (\ref{extSch}), as long as the interior solution (\ref{psoln})-(\ref{fsoln}) 
remains everywhere regular. Third, the solution is regular everywhere except for at most one $r$ 
in the interior where the denominator in (\ref{psoln})
\be
D \equiv 3 \sqrt{1 - H^2 R^2} - \sqrt{1 - H^2 r^2}
\label{denom}
\ee
may vanish in the interval $r \in [0,R]$. Fourth, and most importantly, since $f = \frac{1}{4}D^2$, if $D=0$ 
the pressure $p(r)$ diverges at the {\it same} value as that at which $f(r)$ vanishes. Otherwise
$f(r) > 0$ and the interior solution is regular everywhere else with no horizon.

\begin{figure}[ht] 
\begin{center}
\includegraphics[height=6.7cm, trim=5cm 1cm 1cm 0.5cm, clip=false]{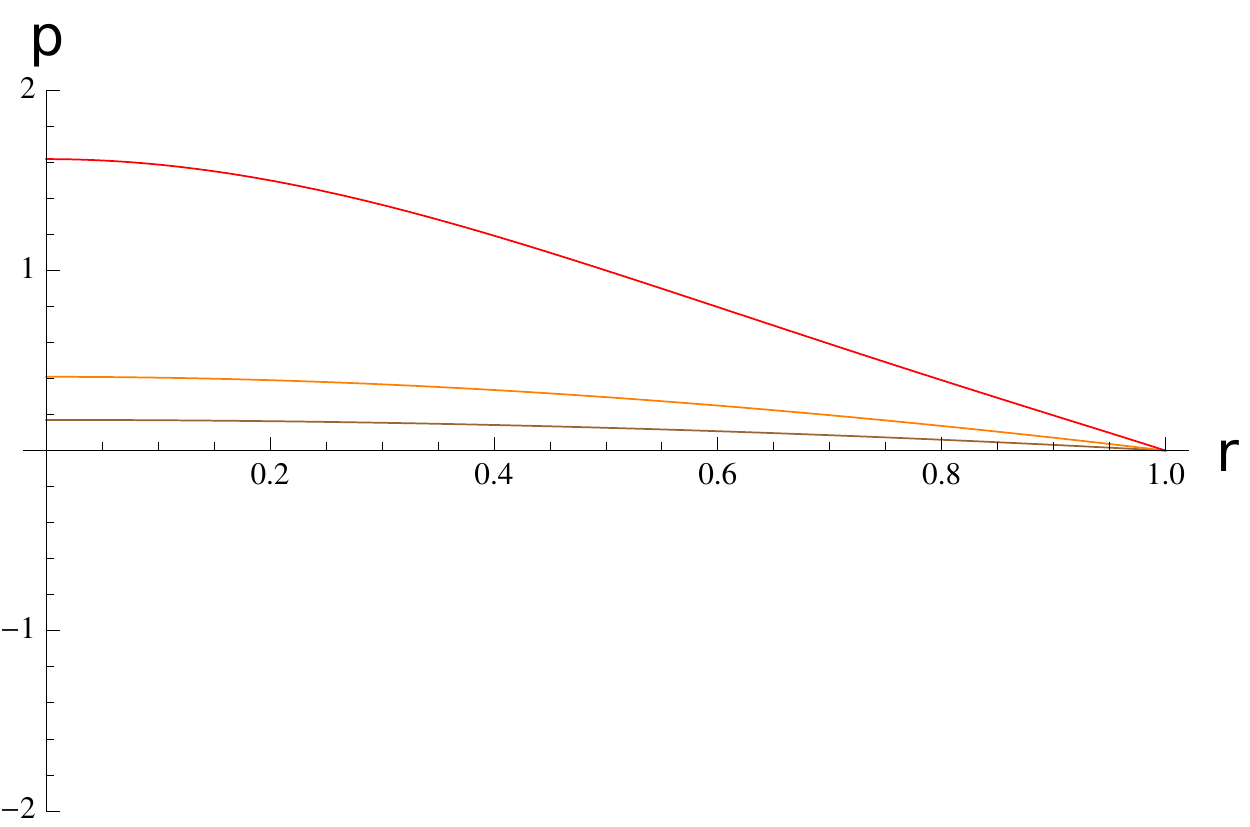}\\
\includegraphics[height=6.7cm, trim=1.5cm 0cm 4cm 1cm, clip=false]{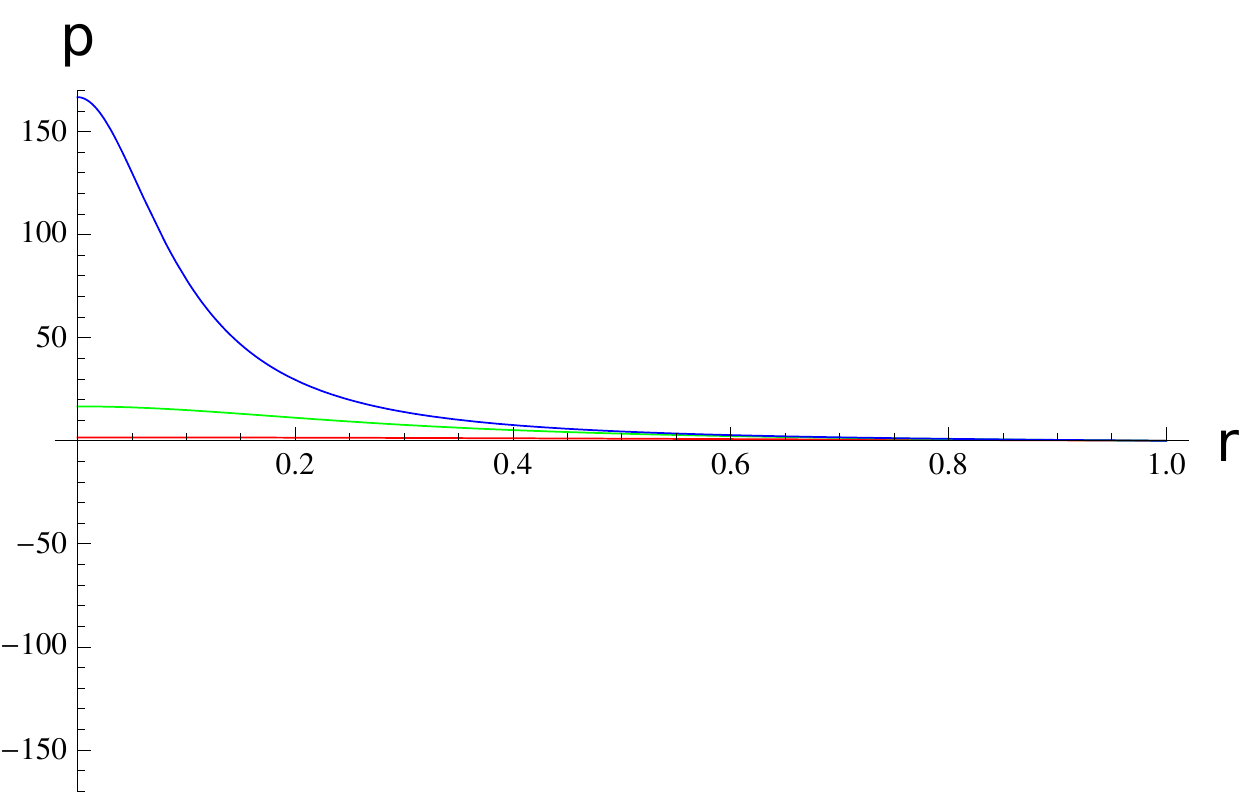}
\end{center}
\caption{Pressure (in units of $\bar\rho$) as a function of $r$ (in units of $R$) of the interior Schwarzschild
solution for various values of the parameter $R/R_s > 9/8 = 1.125$. The upper plot shows $p(r)$
for the values $R/R_s = 2.50, 1.67, 1.25$ (brown, orange, red curves) respectively. The lower plot 
shows $p(r)$ for the values $R/R_s = 1.250, 1.136, 1.126$ (red, green, blue curves) respectively. 
Note the change of vertical scale in the latter plot (the red curves are the same in each) and the rapid 
increase of the central pressure $p(0)$ as $R/R_s$ approaches the Buchdahl bound $1.125$.}
\vspace{-4mm}
\label{Fig:Pressurereg} 
\end{figure}

\begin{figure}[ht] 
\begin{center}
\includegraphics[height=6.7cm, trim=2cm 7mm 2cm 0cm, clip=false]{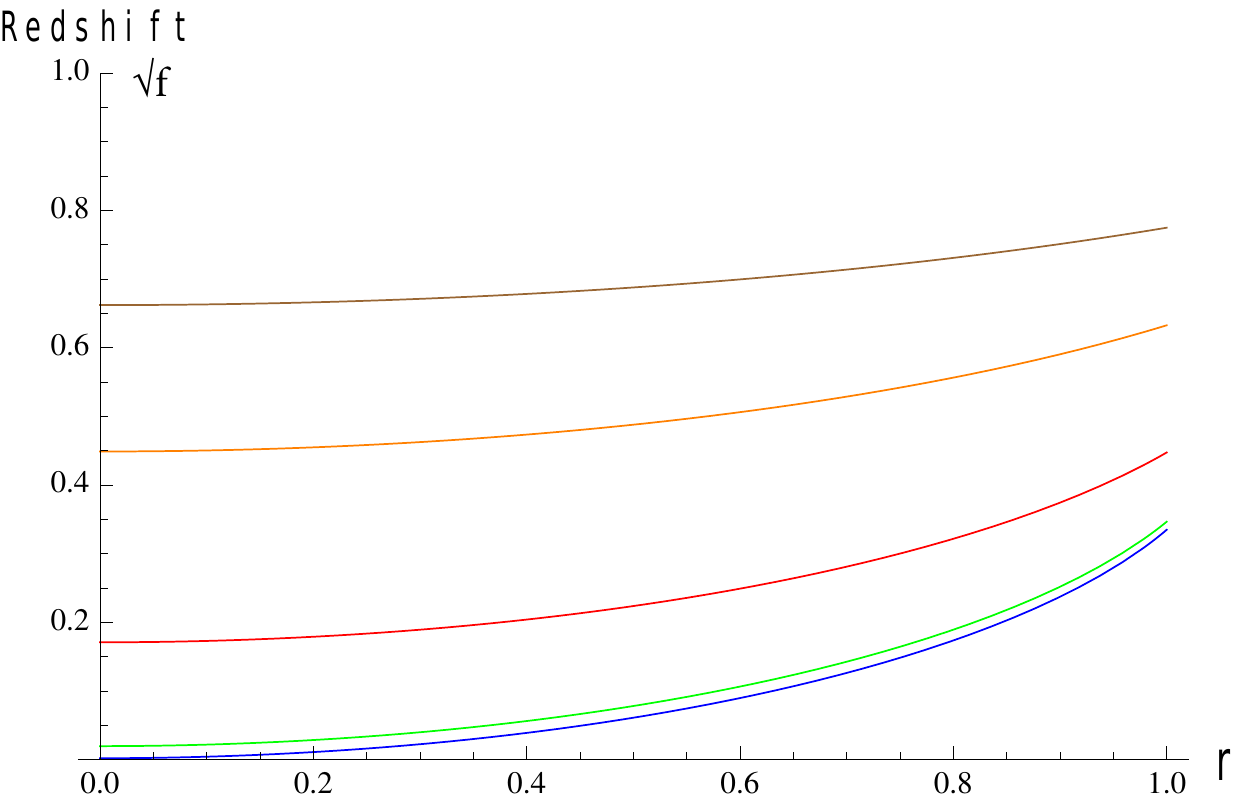}
\end{center}
\caption{The redshift factor $f^{\frac{1}{2}}$ as a function of $r$ (in units of $R$) of the interior Schwarzschild
solution for the same values of the parameter $R/R_s > 9/8 = 1.125$ as in Figs. \ref{Fig:Pressurereg}.
The brown, orange, red, green and blue curves are for the values $R/R_s = 2.50, 1.67, 1.25, 1.136, 1.126$
respectively. Note the approach of $f^{\frac{1}{2}}$ to zero at $r=0$ as $R/R_s$ approaches the Buchdahl 
bound $1.125$.}
\vspace{-3mm}
\label{Fig:Redshiftreg} 
\end{figure}

The solution of $D=0$ is given by $r=R_0$ where
\be
3 \,\sqrt{1 - H^2 R^2} = \sqrt{1 - H^2 R_0^{\ 2}}\qquad {\rm or}\qquad R_0 = 3R\,  \sqrt{1 - \frac{8}{9}\frac{R\,}{R_s}}
\label{R0div}
\ee
which is pure imaginary if $R >\frac{9}{8}R_s$. Hence in this case $D >0$ and the solution (\ref{psoln}) 
remains finite everywhere on the real axis $[0,R]$ in the interior of the star. The pressure (\ref{psoln}) 
is everywhere positive and monotonically decreasing outward from its maximum at $r=0$,
and $f(r)$ remains strictly positive everywhere in this case. The positive regular functions $p(r)$ and 
$f^{\frac{1}{2}}(r)$ (called the {\it redshift factor}) are plotted in Figs. \ref{Fig:Pressurereg} and \ref{Fig:Redshiftreg} 
for several values of $H^2R^2 = R/R_s > \frac{9}{8}= 1.125$. 

Now as $R\rightarrow \frac{9}{8}\,R_s$ from above, the zero of the denominator $D$ (\ref{R0div}) approaches
the real axis at $R_0 =0$ and a divergence of the central pressure $p(0) \rightarrow \infty$ 
appears with $f(0) \rightarrow 0$ at this same point. Hence at the critical value $R = \frac{9}{8}R_s$, 
the constant  density solution (\ref{psoln}) with $p_{\perp} = p$ everywhere finite strictly ceases to exist. 
We analyze this divergence of the pressure in the next several sections. If nevertheless we consider 
(\ref{psoln})-(\ref{fsoln}) for $R_s < R < \frac{9}{8}R_s$ the zero of $D$ at $R_0$ given by (\ref{R0div}) 
moves outward from the origin to finite values of $0 < R_0 < R$. Then (\ref{psoln}) shows that a new 
regular interior solution opens up for $0 \le r < R_0$ where $D <0$ and $p(r)<0$, while $f(r)$ is again 
positive. 

\begin{figure}[ht] 
\begin{center}
\includegraphics[height=7.5cm, trim=6cm 0cm 2cm 0cm, clip=false]{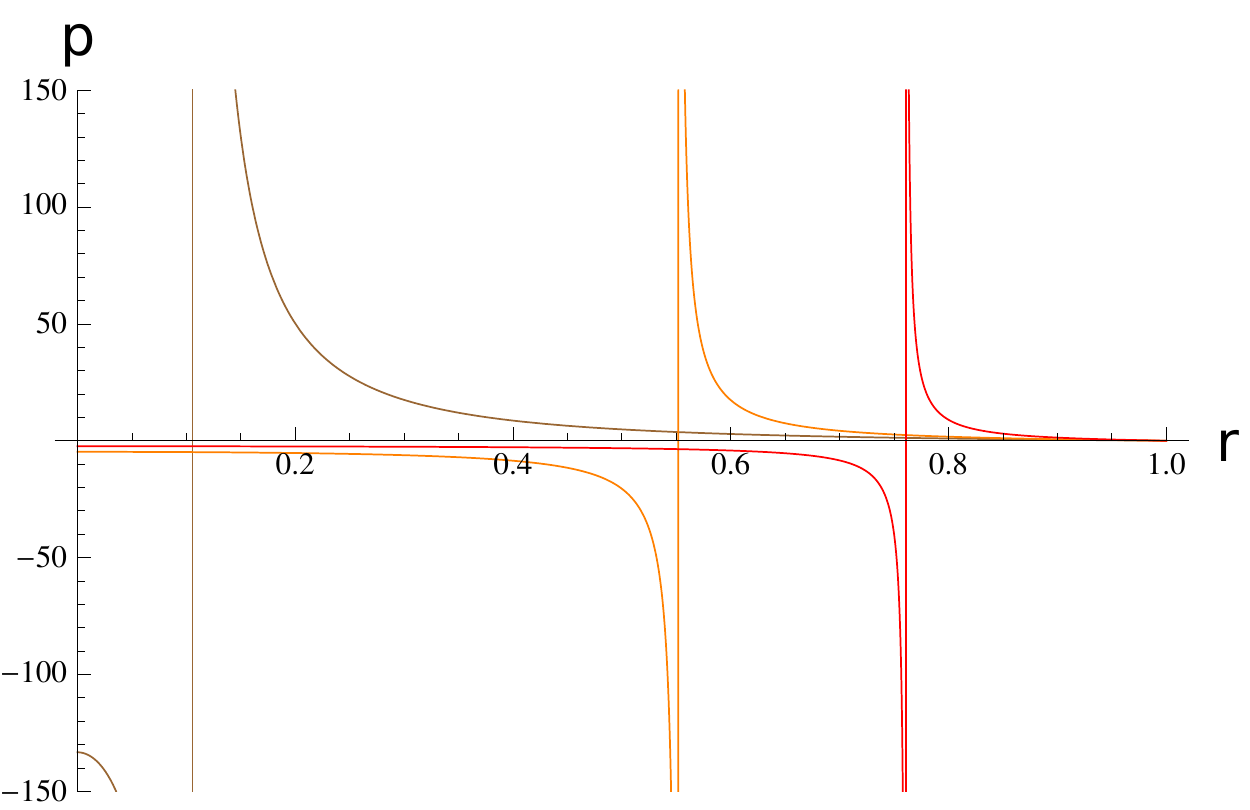}\\
\includegraphics[height=6cm, trim=2cm 0cm 6cm 1cm, clip=false]{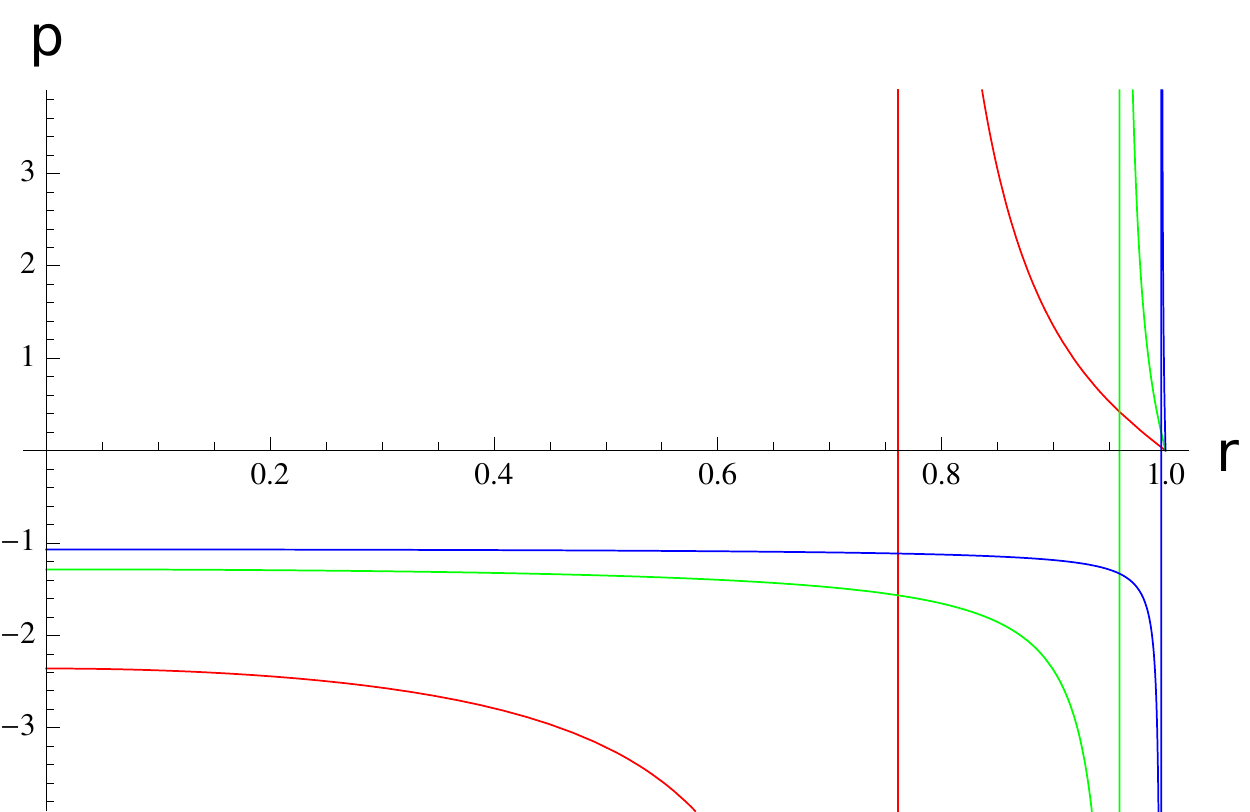}
\end{center}
\caption{Pressure (in units of $\bar\rho$) as a function of $r$ (in units of $R$) of the interior Schwarzschild
solution for various values of the parameter $R/R_s < 9/8 = 1.125$. The upper plot shows $p(r)$
for the values $R/R_s = 1.124, 1.087, 1.053$ (brown, orange, red curves), where the
divergence in the pressure occurs at $R_0/R = 0.106, 0.552, 0.761$ respectively. The lower plot 
shows $p(r)$ for the values $R/R_s = 1.053, 1.010, 1.001$ (red, green, blue curves), where the
divergence in the pressure occurs at $R_0/R = 0.761, 0.959, 0.996$ respectively. For $r <R_0$
the pressure is negative. Note the change of vertical scale in the plots (the red curves are the 
same in each) and the approach of the negative interior pressure $p \rightarrow -\bar\rho$
as $R$ approaches the  Schwarzschild radius $R_s$ from above and $R_0$ approaches 
$R_s$ from below.}
\vspace{-3mm}
\label{Fig:Pressuresing} 
\end{figure}

\begin{figure}[ht] 
\begin{center}
\includegraphics[height=6.7cm, trim=2cm 7mm 2cm 0cm, clip=false]{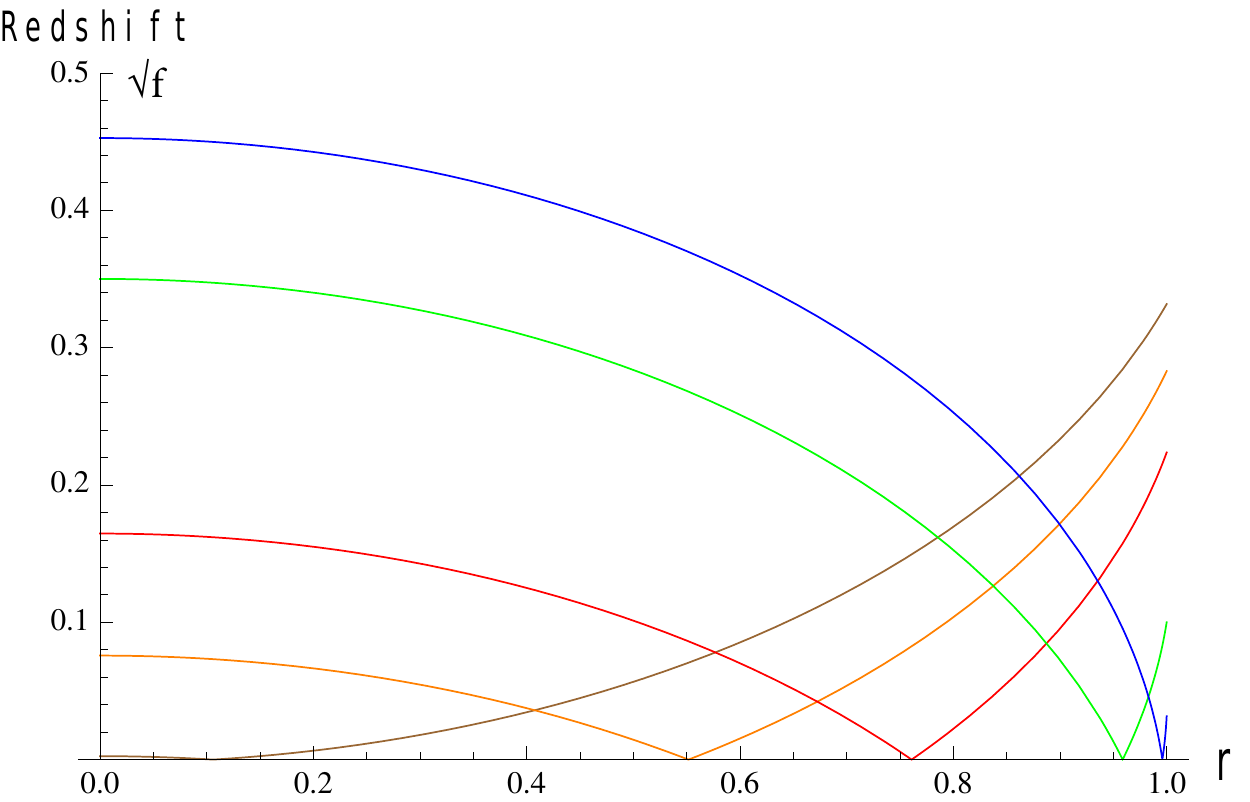}
\end{center}
\caption{The redshift factor $f^{\frac{1}{2}}$ as a function of $r$ (in units of $R$) of the interior Schwarzschild
solution for the same values of the parameter $R/R_s < 9/8 = 1.125$ as in Figs. \ref{Fig:Pressuresing}.
The brown, orange, red, green and blue curves are for the values $R/R_s = 1.124, 1.087, 1.053,1.010, 1.001$
respectively. Note the approach of the zero of $f^{\frac{1}{2}}$ at $R_0$ towards $R$ from below as $R$ 
approaches the Schwarzschild radius $R_s$ from above.}
\label{Fig:Redshiftsing} 
\end{figure}

As the star is compressed further and its radius approaches the Schwarzschild radius 
$R \rightarrow R_s^+$ from outside, (\ref{R0div}) shows that the radius of the sphere where the pressure 
diverges and $f(R_0) = 0$ moves from the origin to the outer edge of the star, {\it i.e.} $R_0 \rightarrow R_s^-$, 
and in that limit the interior solution with negative pressure comes to encompass the entire interior 
region $0 \le r < R$, excluding only the outer boundary at $R= R_s$. Finally, and most remarkably of all, 
since in this limit $H^2 R^2 = R_s/R \rightarrow 1$ inspection of (\ref{psoln}) shows that
the entire interior solution then has {\it constant negative pressure}
\be
p = - \bar \rho\,,\qquad {\rm for} \qquad r < R=R_0 = R_s = 2GM
\ee
with
\be
f(r) = \tfrac{1}{4}\, (1 - H^2 r^2) = \tfrac{1}{4}\, h(r) = \frac{1}{4}\, \left(1 - \frac{r^2}{R_s^2}\right), \qquad H = \frac{1}{R_s}
\label{fhint} 
\ee
corresponding to a patch of pure de Sitter space in static coordinates, although one in which
$g_{tt}$ is $\frac{1}{4}$ its usual value, so that the passage of time in the interior is modified from what
would be expected in the usual static coordinates of de Sitter space. 
Typical profiles of the pressure $p(r)$ and $f^{\frac{1}{2}}$ for values of the radius in the range 
$R_s < R < \frac{9}{8}R_s$ and the approach to $R_s$ are shown in Figs. \ref{Fig:Pressuresing} 
and \ref{Fig:Redshiftsing}.

\begin{figure}[ht] 
\begin{center}
\includegraphics[height=6.7cm, trim=2cm 7mm 2cm 0cm, clip=false]{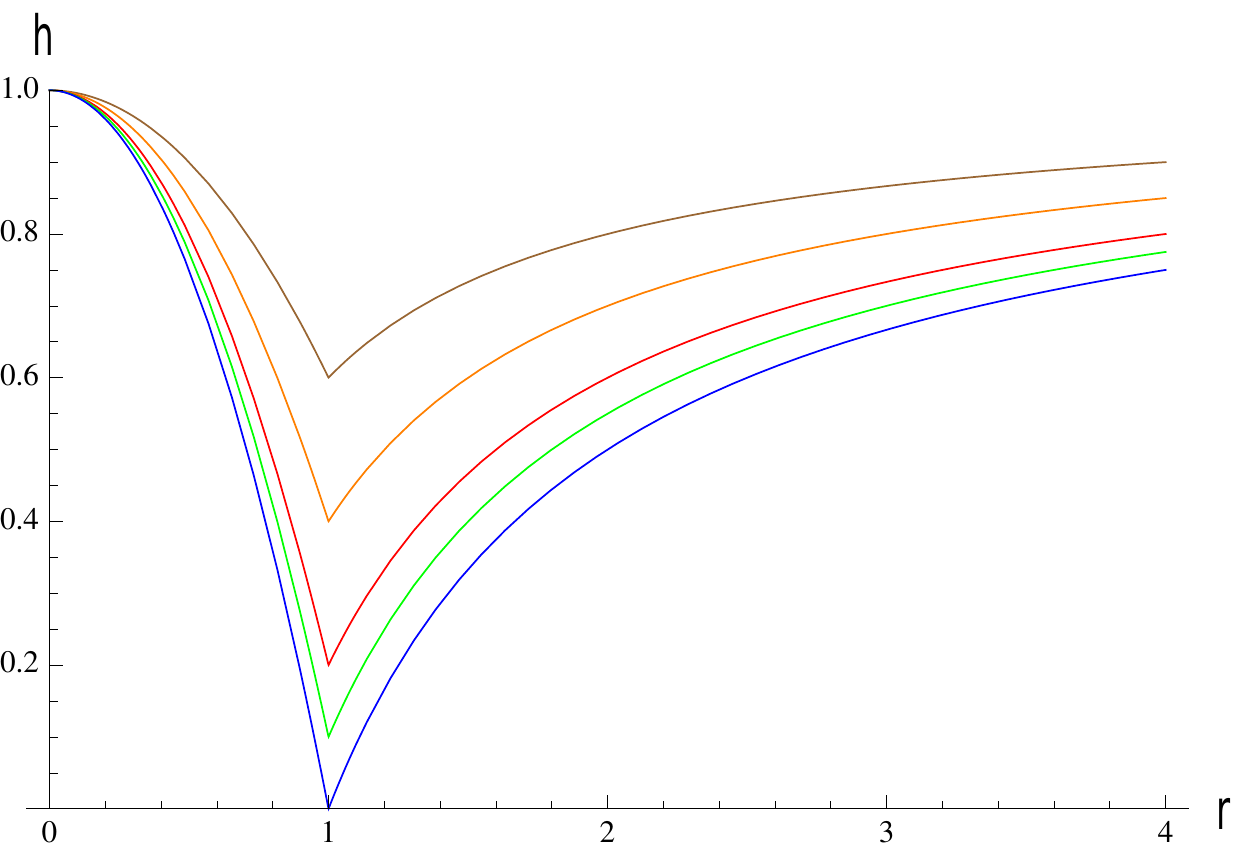}
\end{center}
\caption{The metric function $h$ as a function of $r$ (in units of $R$) of the interior and exterior Schwarzschild 
solution for the values of the parameter $R/R_s  = 2.500, 1.667, 1.250,1.111, 1.000$  (brown, orange, red, green 
and blue curves) respectively. The minimum of $h$ approaches zero at $r=R$, as $R$ as $R$ approaches the 
Schwarzschild radius $R_s$ from above. Note the cusp-like behavior in both this and the previous figure.}
\label{Fig:hvarious} 
\vspace{-3mm}
\end{figure}

The second metric function $h(r)$ for the same values of the radius in the range $R_s < R < \frac{9}{8}R_s$ 
and the approach to $R_s$ are shown in Fig. \ref{Fig:hvarious}. When $R=R_s=R_0$, the exterior region $r > R_s$ 
is the usual vacuum Schwarzschild solution (\ref{extSch}) with an infinitely thin shell discontinuity and jump 
of the pressure at $r=R_s$, where the horizons $f=h=0$ of the interior de Sitter and exterior Schwarzschild 
solutions coincide. Note that both $f(r)$ and $h(r)$ are non-analytic in $r$ at $R_0$, having a cusp there,
and being positive on either side of $r=R_0 \rightarrow R_s$, although $r=R_s$ is a null surface,
there is no interior trapped surface, and no causal event horizon.

Because of the pressure divergence at $R_0$ which first appears at the origin when $R = \frac{9}{8} R_s$
the interior Schwarzschild negative pressure solution has been little studied or remarked upon in the
literature \cite{CatFabVis}. We show in the next two sections by use of the covariant Komar mass-energy 
integral, containing in its integration measure the redshift factor $f^{\frac{1}{2}}$ which vanishes at the 
same radius $R_0$,  that the pressure singularity is integrable, and requires relaxation of the
$p_{\perp} = p$ perfect fluid isotropy assumption. When $R < \frac{9}{8} R_s$ the non-isotropic pressure 
$p_{\perp} \neq p$ develops a $\delta$-function distribution at $r=R_0$ whose coefficient corresponds to 
a non-zero surface tension,  which is therefore inherent in the Schwarzschild interior solution. Then the regular 
negative pressure solution for $R \le \frac{9}{8} R_s$ and $r <R_0$ is recognized as a perfectly viable solution, 
with a surface boundary layer separating the region of negative from positive pressures. In the limit 
$R\rightarrow R_s^+$, $R_0\rightarrow R_s^-$, it becomes a non-singular classical alternative to a black 
hole for the fully collapsed state. In fact, in this limit the constant density solution found by Schwarzschild 
nearly a century ago becomes essentially the gravitational condensate star solution with a surface boundary 
layer at $R_s$, proposed in \cite{gstar,PNAS}.

\section{Mass-Energy Flux and Surface Gravity}
\label{Sec:Mass}

Properly interpreting the pressure singularity of the constant density interior Schwarzschild solution requires some
background and formalism which we provide in these next three sections, starting in this section with
the Komar mass-energy formula, energy flux and surface gravity.

In electrodynamics Maxwell's eqs. $\nabla_{\nu}F^{\mu\nu} = 4 \pi J^{\mu}$ together with Stokes' theorem allows 
integration of the anti-symmetric field strength tensor on a two-surface $S$ with directed surface element $d\Sigma_{\mu\nu}$,
\be
 \int_S  F^{\mu\nu}\, d\Sigma_{\mu\nu} = \int_V  \nabla_{\!\nu}F^{\mu\nu}\, d\Sigma_{\mu} 
=4 \pi  \int_V J^{\mu}\, d\Sigma_{\mu} =4 \pi  \int_V\, J^0 dV = 4 \pi\, Q
\label{charge}
\ee
thereby expressing the flux integral of the field through the surface in terms of a volume integral for the total conserved charge $Q$.
Here $d\Sigma_{\mu} = \delta_{\ \mu}^0\, dV$ is a spacelike three-volume element of flat space with timelike normal.

In General Relativity the analogous covariant volume integral over the matter distribution which yields the total mass-energy 
of an isolated stationary system in a general asymptotically flat spacetime (not necessarily spherically symmetric)
was given first by Tolman \cite{Tolm} and re-derived more geometrically by Komar \cite{Komar}.  The latter derivation 
relies upon the fact that a stationary spacetime is invariantly characterized as one admitting a timelike Killing vector field 
$K^{\mu}$, satisfying the Killing equation
\be
\nabla^{\mu}K^{\nu} + \nabla^{\nu}K^{\mu} = 0
\label{Killvec}
\ee
so that $\nabla^{\mu}K^{\nu} = \nabla^{[\mu}K^{\nu ]}$ is anti-symmetric in its indices. Thus just as $F^{\mu\nu}$ in (\ref{charge}),  
this anti-symmetric tensor may be integrated over a two-surface and Stokes'  theorem applied to yield
\be
\int_S \nabla^{\mu}\!K^{\nu}\, d\Sigma_{\mu\nu} = \!\int_V \nabla_{\nu}\nabla^{\mu}\!K^{\nu} \, d\Sigma_{\mu}
=  \!\int_V R^{\mu}_{\ \lambda}K^{\lambda}\, d\Sigma_{\mu} =  
4 \pi G  \int_V \big(2\, T^{\mu}_{\ \lambda} - T\, \delta^{\mu}_{\ \lambda}\big)\,K^{\lambda}\, d\Sigma_{\mu} 
\label{intK}
\ee
where the commutator of two covariant derivatives in terms of the curvature tensor is used, together with
$\nabla_{\nu}K^{\nu} = 0$, and the Einstein eqs. $R^{\mu}_{\ \lambda} = 4 \pi G\, 
(2\, T^{\mu}_{\ \lambda} - T\, \delta^{\mu}_{\ \lambda})$ with $T \equiv T^{\mu}_{\ \mu}$. 

It is natural to introduce the time coordinate $t$ with respect to which the spacetime is stationary, as in (\ref{Killt})-(\ref{Killnorm}). 
If the integration in (\ref{intK}) is taken to be a spacelike hypersurface at constant $x^0=t$, then in curved space
\vspace{-5mm}
\be
d\Sigma_{\mu} = e^0_{\ \mu}\,  dV
\ee
where $e^0_{\ \mu}$ is the vierbein in the orthonormal basis, $g^{\mu\nu} e^a_{\ \mu}e^b_{\ \nu} = \eta^{ab} =$diag $(-1,1,1, 1)$, 
and $dV$ is the three-volume element in the induced three-metric of the hypersurface. In these natural coordinates suited to 
stationarity, the inverse vierbein $E^{\ \mu}_0= f^{-\frac{1}{2}} K^{\mu} =  u^{\mu}$, where $u^{\mu}$ is the four-velocity of a 
particle at rest with respect to the time $t$ and $f \equiv -K_{\mu}K^{\mu} = -g_{tt}$ is a spacetime scalar. Thus 
\be
e^0_{\ \mu} = - f^{-\frac{1}{2}} K_{\mu}= -u_{\mu} = f^{\frac{1}{2}}\,\delta^t_{\mu}
\ee 
is the directed normal to the hypersurface and $d\Sigma_{\mu}= -u_{\mu} dV = f^{\frac{1}{2}}\,\delta^t_{\ \mu} dV$.
We note that these formulae making explicit use of the vierbein frame field $e^a_{\ \mu}$ and its inverse $E_a^{\ \mu}$
make sense strictly speaking only if $K^{\mu}$ is timelike and $f  > 0$, the $f\rightarrow 0$ limit
requiring special care.

If furthermore the spatial coordinates $(x^1, x^2, x^3)$ of the hypersurface at constant $t$ are defined so that the 
two-surface in (\ref{intK}) lies at constant $x^1$, then the directed two-surface element may be written
\be
d\Sigma_{\mu\nu} = e^0_{\ [\mu}e^1_{\ \nu]}\, dA
\ee
with $dA$ the area surface element of the induced two-metric of the surface. The (negative of the) integrand of the surface 
integral in (\ref{intK}) is then 
\be
\nabla^{\nu}\!K^{\mu} \,e^0_{\ [\mu}e^1_{\ \nu]} = \nabla^{\nu}\!K^{\mu}\,\left(-f^{-\frac{1}{2}} K_{\mu}\,e^1_{\ \nu}\right) =
\tfrac{1}{2}\, e^1_{\ \nu}  f^{-\frac{1}{2}}\nabla^{\nu} f = E^{\ \nu}_{1}\, \partial_{\nu} f^{\frac{1}{2}} \equiv \kappa_f\,.
\label{surfgrav}
\ee
This quantity is related to the proper four-acceleration of a particle at rest, namely
\pagebreak
\be
a^{\nu} \equiv \frac{du^{\nu} }{d \tau\,}= u^{\mu} \nabla_{\mu} u^{\nu} = \frac{1}{f}\,  K^{\mu}\nabla_{\mu} K^{\nu} = 
- \frac{1}{f}\,  K^{\mu}\nabla^{\nu} K_{\mu} = \frac{1}{2f}\, \nabla^{\nu}\! f
\label{accel}
\ee
where $K^{\nu}\nabla_{\nu} f = \partial_t f = 0$ has been used. Multiplying (\ref{accel}) by the redshift factor $f^{\frac{1}{2}}$ 
converts the acceleration with respect proper time $\tau$ to that with respect to the stationary coordinate time $t$. Taking 
the projection of  $f^{\frac{1}{2}}a^{\nu}$ lying within the spatial hypersurface in the direction normal to the two-surface 
by contracting with $e^1_{\ \nu}$ then gives (\ref{surfgrav}). Thus $\kappa_f$ is the four-acceleration of the worldline of a particle 
at rest with respect to the time $t$ projected onto the normal to the surface $x^1=$ constant.

For a three-volume $V$ enclosed by an outer and inner two-surface $\partial V_+$ and $\partial V_-$ respectively, (\ref{intK})
then becomes
\vspace{-3mm}
\be 
\frac{1}{4 \pi G} \int_{\partial V_+}\!\!\! \kappa_f\, dA = \!\int_V \big(2\, T^{\mu}_{\ \lambda} - T\, \delta^{\mu}_{\ \lambda}\big) \,
K^{\lambda} u_{\mu} \, dV\ +\ \frac{1}{4 \pi G} \int_{\partial V_-}\!\!\! \kappa_f\, dA 
\label{intVA}
\ee
after dividing by $4\pi G$ and rearranging. This shows that if the volume $V$ contains no matter, the areal surface integral 
of $\kappa_f$ is independent of the surface $\partial V$ chosen, and $\kappa_f$ is proportional to the conserved mass-energy 
flux through the surface, analogous to the electric flux normal to the surface of Gauss Law (\ref{charge}) in electromagnetism. 
The coefficient $1/4 \pi G$ has been fixed so that if the surface integral over $\partial V_+$ is taken outside the matter distribution, 
it evaluates to the total mass $M$ in the case of an asymptotically flat spacetime. Thus in the asymptotically flat case 
(\ref{intVA}) becomes
\be
M = \int_V\! \big(2\, T^{\mu}_{\ \lambda} - T\, \delta^{\mu}_{\ \lambda}\big)\, K^{\lambda}u_{\mu}\, dV
\ +\ \frac{1}{4 \pi G} \int_{\partial V_-}\!\!\! \kappa_f\, dA 
\label{KomarM}
\ee
expressing the total mass-energy of the system $M$ in terms of a three-volume integral of the matter stress-energy,
plus a possible surface flux contribution from the inner two-surface. The redshift factor $f^{\frac{1}{2}}$ is then the
gravitational redhsift relative to the asymptotically flat region where $f=1$. 

If the volume integral over $V$ can be extended to a complete Cauchy surface without an inner boundary, such as in the 
case of a (non-singular) star, then the last surface integral contribution at $\partial V_-$ in (\ref{KomarM}) is absent, 
and (\ref{KomarM}) gives the total mass of an arbitrary isolated stationary system in asymptotically flat spacetime, 
in terms of a certain volume integral of its stress-energy components.

The relation (\ref{KomarM}) applied to vacuum solutions of the Einstein eqs. such as the Kerr rotating 
black hole family of solutions, breaks down at the ergosphere boundary where $f=0$ and it is no longer possible for a 
particle to remain stationary. In order to extend (\ref{KomarM}) within the ergosphere the usual route is to express 
$K^{\mu} = \ell^{\mu} - \Omega_H \tilde K^{\mu}$ as a linear combination of the rotational Killing vector 
$\tilde K^{\mu} \partial_{\mu} = \partial/\partial \phi$ and another vector $\ell^{\mu}$ 
which remains timelike outside the black hole horizon, becoming null there. A modified surface gravity 
can be defined then with respect to this vector, with its corresponding redshift factor $(-\ell_{\mu}\ell^{\mu})^{\frac{1}{2}}$ 
remaining finite and well-defined down to the black hole horizon, where it is becomes a constant, $\kappa_H$. The contribution 
$\int_S \nabla^{\nu} \tilde K^{\mu} d\Sigma_{\mu\nu}$ can be evaluated in terms of the angular momentum $J_H$ 
of the black hole. In this way one obtains Smarr's integral mass formula \cite{Smar,BarCarHaw} 
\vspace{-3mm}
\be
M = \frac{\kappa_H}{4 \pi G}\  A_H + 2\, \Omega_H J_H 
\label{SmarM}
\ee
for a rotating black hole, with $A_H$ the area of the Kerr black hole horizon and $\Omega_H$ its angular velocity of rotation.
The differential form of this relation \cite{Smar,BarCarHaw}
\be
dM = \frac{\kappa_H}{8\pi G} \, dA_H + \Omega_H \, dJ_H
\label{dM}
\ee
expresses the change of total energy of the system in terms of the change of angular momentum and change
of surface area of the horizon, and has been called the First Law of Black Hole Mechanics \cite{BarCarHaw}. 

The differential form (\ref{dM}) suggests that the coefficient
\be
\left(\frac{\partial\, M}{\partial A_H\!}\right)_{\!\!J_H}  = \frac{\kappa_H}{8\pi G} 
\label{dMdA}
\ee
could perhaps be viewed as the {\it surface tension} of the classical black hole horizon \cite{Smar}. Yet this interpretation
is problematic in black hole physics, since if globally extended within its event horizon by analytic 
continuation, it is implicitly assumed that a black hole has no stress-energy whatsoever localized on the horizon. 
Thus it is not clear to what surface energy or surface effect on the horizon the `surface tension' (\ref{dMdA}) could 
possibly be associated. We shall see that the identification of surface gravity with surface tension (actually the
difference of surface gravities between exterior and interior) is possible only when the regular interior solution 
such as the constant density Schwarzschild interior solution is known, and when this solution differs fundamentally
from that obtained by analytic continuation of the exterior vacuum Schwarzschild solution.

By covariant differentiation of the Killing eq. (\ref{Killvec}), the integrand appearing in
(\ref{KomarM}) may be written also in the local form
\be
-\sq K^{\mu} = 8 \pi G\, \Big( T^{\mu}_{\ \nu} - \tfrac{1}{2}\,\delta^{\mu}_{\ \nu}\, T\Big)K^{\nu}
\label{locMass}
\ee
prior to any integration. For the general spherically symmetric static form of the metric (\ref{sphsta})
and stress-energy tensor (\ref{Tgeneral}) this eq. takes the form
\be
\frac{d}{dr} \left[r^2 \sqrt{\frac{h}{f}}\ \frac{df}{dr}\right] = 8 \pi G\, \sqrt{\frac{f}{h}} \ r^2\, \big(\rho + p + 2 p_{\perp}\big)
\label{locr}
\ee 
in Schwarzschild coordinates  (\ref{sphsta}). In terms of the surface gravity $\kappa_f = \kappa$ of (\ref{surfgrav}),
(dropping the subscript $f$ henceforth)
\be
\kappa (r) = \frac{1}{2} \sqrt{\frac{h}{f}}\ \frac{df}{dr} 
\label{surfgravfh}
\ee
so that (\ref{locr}) can be written in the form
\be
\frac{1}{G}\frac{d}{dr} \left(r^2 \kappa(r) \right) = 4 \pi\, \sqrt{\frac{f}{h}} \ r^2\, \big(\rho + p + 2 p_{\perp}\big)
\label{Mintegrand}
\ee
justifying the interpretation of $\kappa/G$ as a mass-energy flux, analogous to the electric flux
in Gauss' Law. Indeed since in the exterior vacuum Schwarzschild solution (\ref{extSch})
\be
\kappa_{ext}(r) = \frac{GM}{r^2}
\ee
the upper limit of the integration of (\ref{Mintegrand}) outside the matter distribution gives the total mass 
$M$.

\section{Surface Energy and Surface Tension}
\label{Sec:Tension}

The local form (\ref{Mintegrand}) in the spherically symmetric case is the convenient starting point 
to analyze the singular behavior of the pressure (\ref{psoln}) of the constant density solution at $r=R_0$. 
We treat the case of general $0\le R_0 < R$ and consider the limit $R \rightarrow R_s^+$ when 
$R_0 \rightarrow R_s^-$ at the end. If one substitutes the constant density interior Schwarzschild 
solution assuming $p_{\perp}(r) = p(r)$ given by (\ref{psoln}) and 
\vspace{-5mm}
\be
\sqrt f = \tfrac{1}{2} \,|D|
\label{fD}
\ee
from (\ref{fsoln}) into the right side of (\ref{Mintegrand}) one finds 
\pagebreak
\vspace{-3mm}
\be
4 \pi\, \sqrt{\frac{f}{h}} \ r^2\, \big(\rho + 3p\big) = 4 \pi \, r^2\,\bar\rho \ {\rm sgn}\, (D) \,,\qquad r\neq R_0
\label{rhslocr}
\ee
so that the divergence at $r=R_0$ apparently cancels, leaving only a sign function discontinuity
\vspace{-3mm}
\be
{\rm sgn}\, (D) = {\rm sgn}\, (r- R_0) = \left\{\begin{array}{l} -1, \quad r< R_0\\ +1, \quad r> R_0\end{array} \right. 
\ee
at $r=R_0$. The cancelation of the divergence in (\ref{rhslocr}) indicates that the pressure
singularity is an integrable one with respect to the proper measure in the mass-energy integral 
(\ref{KomarM}) or (\ref{Mintegrand}). However, since from (\ref{denom}) and (\ref{fD}), with $p \propto D^{-1}$, 
and $f \rightarrow 0$, we are dealing with singular distributions rather than smooth functions, and this conclusion 
is unreliable at the singular point itself, potentially missing a local integrable distribution with support only at $r=R_0$. 

That such a $\delta$-function is indeed present at $r=R_0$ is verified by examining the left side 
of (\ref{Mintegrand}), which is generally valid for all static, spherically symmetric spacetimes. 
Substituting the interior solution (\ref{mhr}), (\ref{fsoln}) into the quantity to be
differentiated on the left side of (\ref{Mintegrand}) we obtain
\be
\frac{r^2}{G} \, \kappa (r) = \frac{r^2}{2G}\,  \sqrt{\frac{h}{f}}\ \frac{df}{dr} = \frac{4 \pi}{3}\,\bar \rho \,  r^3 \ {\rm sgn} \, (D)\,.
\ee
The derivative $d/dr$ then produces a $\delta$-function contribution by differentiation of the sign
function discontinuity,
\vspace{-5mm}
\be
\frac{d}{dr}\, {\rm sgn}\, (D) = \frac{dD}{dr} \, \frac{d}{dD}\, {\rm sgn}\, (D) = 2\, \frac{dD}{dr} \,\delta (D) = 2\, \delta (r-R_0)
\ee
since $\displaystyle\Big( \frac{dD}{dr}\Big)_{R_0} = \bigg\vert \frac{dD}{dr}\bigg\vert_{R_0}$ is an even function at $r=R_0$. 
Thus the left side of (\ref{Mintegrand}) is {\it in toto}
\be
\frac{1}{G} \frac{d}{dr} \Big(r^2 \kappa(r)\Big) = 4 \pi  \, r^2 \bar\rho \ {\rm sgn} (D) 
+ \frac{8 \pi }{3} \,R_0^{\ 3}\, \bar \rho\ \delta (r-R_0)
\label{lhslocr}
\ee
with a well-defined local $\delta$-function contribution having support only at $r=R_0$ in addition to the finite contribution
given previously by (\ref{rhslocr}). Comparing the general (\ref{Mintegrand}) with (\ref{rhslocr}) and (\ref{lhslocr}),
the $\delta$-function distribution must be attributed to the difference
\be
8 \pi \sqrt{\frac{f}{h}} \ r^2\,  (p_{\perp} - p) = \frac{8 \pi }{3} \, \bar \rho\, R_0^{\ 3}\,\delta (r-R_0)
\label{pdiff}
\ee
and hence the breakdown of the isotropic pressure assumption $p_{\perp} = p$ at the singular radius $r=R_0$.

This interpretation of the $\delta$-function contribution may be confirmed from the pressure balance eq. (\ref{cons}) 
expressed in the form
\vspace{-6mm}
\be
r \frac{d}{dr}\, \big[(p+ \bar\rho) f^{\frac{1}{2}}\big] = 2\, (p_{\perp} -p) f^{\frac{1}{2}}
\label{pressjump}
\ee
for constant density  $\rho = \bar\rho$. Substituting the solution (\ref{psoln})-(\ref{fsoln}) in the left side 
of this relation gives
\be
r \frac{d}{dr}\Big[\bar\rho \sqrt{1-H^2R^2}\,{\rm sgn}\,(D)\Big] = 2\bar\rho\, R_0\, \sqrt{1-H^2R^2}\ \delta(r-R_0) =
\frac{2}{3}\,\bar\rho\, R_0\, \sqrt{1-H^2R_0^2}\ \delta(r-R_0)\,.
\ee
Multiplying by $4\pi r^2h^{-\frac{1}{2}}$, evaluated at $r=R_0$ then yields (\ref{pdiff}) once again.

Thus the integrand of the right side of the Komar mass-energy (\ref{Mintegrand}) may be written in the form
\be
4 \pi \sqrt{\frac{f}{h}} \ r^2\, \Big[\rho + 3p + 2\,(p_{\perp}- p)\Big] = 4 \pi \, r^2\,\bar\rho \, {\rm sgn}\, (D)
+ \frac{8 \pi }{3} \,R_0^{\ 3}\, \bar \rho\ \delta (r-R_0)
\label{rhsMint}
\ee
in agreement with (\ref{lhslocr}). In Appendix \ref{Sec:DeltaDistrib} we show how the $\delta$-function can be obtained 
also on the right side of (\ref{locr}) of (\ref{Mintegrand}) by a careful regularization of the singularity at $r=R_0$ and taking 
the limit properly of removing the regulator. It is this $\delta$-function contribution, totally integrable within the Komar mass 
formula (\ref{Mintegrand}), but breaking the assumed isotropic perfect fluid condition $p_{\perp} = p$ at $r= R_0$, and the fact 
that the radial pressure is a Principal Part distribution which is also integrable, that allows the interior Schwarzschild solution to 
be interpreted in physical terms, despite failing to satisfy the Buchdahl bound for $R \le \frac{9}{8}R_s$.

From (\ref{lhslocr}) or (\ref{rhsMint}) we see that the $\delta$-function contribution gives a surface energy contribution
\vspace{-2mm}
\be
E_s=  \frac{8 \pi}{3} \,\bar \rho\, R_0^{\ 3} = 2M \left(\frac{R_0}{R}\right)^3
\label{Esurf}
\ee
to the total Komar mass-energy integral (\ref{KomarM}). This is attributable to the discontinuous change of sign of
the surface gravity (\ref{surfgrav}) as $R_0$ is approached from above and below, namely
\vspace{-1mm}
\be
\kappa_{\pm} \equiv \lim_{r \rightarrow R_0^{\pm}}  \kappa(r) = \pm \frac{4 \pi G}{3}\, \bar\rho\,R_0 
= \pm \frac{GMR_0}{R^3}
\label{kappm}
\ee
so that the discontinuity in the surface gravities is
\vspace{-1mm}
\be
\Delta \kappa \equiv \kappa_+ - \kappa_- = \frac{2GMR_0}{R^3} = \frac{R_sR_0}{R^3}
\label{Delkap}
\ee
which leads to a (redshifted) surface tension of the surface at $r=R_0$ of\\
\vspace{-1.1cm}
\be
\tau_s = \frac{E_s}{2 A} = \frac{E_s}{8 \pi R_0^2} = \frac{MR_0}{4 \pi R^3}= \frac{\Delta \kappa}{8 \pi G}\,.
\label{deftau}
\ee
This is a physical surface tension associated with a genuine surface energy and positive integrable
transverse pressure contribution to the integral of the Komar mass formula (\ref{surfgravfh}). 

The Komar mass formula may now be consistently applied to the interior and exterior Schwarzschild
solution throughout the domain $0 \le r < \infty$ on a complete Cauchy hypersurface. Since the integrand 
of (\ref{Mintegrand}) is given by (\ref{lhslocr}) or (\ref{rhsMint}), the volume integral of (\ref{Mintegrand}) 
excluding $r=R_0$ gives
\be
E_v = - \int_0^{R_0^-} 4 \pi r^2 \bar \rho\, dr + \int_{R_0^+}^R  4 \pi r^2 \bar \rho\, dr = M - 2M \left(\frac{R_0}{R}\right)^3
\label{Evol}
\ee
while the $\delta$-function at $r=R_0$ gives the surface contribution (\ref{Esurf}) necessary for the total exterior
Schwarzschild mass $M= E_v + E_s$ to be obtained. Interestingly, the volume bulk contribution to the Komar 
mass-energy in (\ref{lhslocr}) or (\ref{rhsMint}) agrees with the Misner-Sharp mass density $4 \pi r^2 \bar \rho$ 
(\ref{mint}) only in the outer portion $R_0 < r < R$, where sgn $(D) = 1$, which is the entire interior region for 
$R > \frac{9}{8} R_s$, while for $ R_s < R < \frac{9}{8} R_s$ and $0\le r < R_0$ it has the opposite sign.
In the latter case the surface energy contribution (\ref{Esurf}) is necessary for consistency of the Komar mass 
formula with the total integrated Schwarzschild mass $M$ from the sum of (\ref{Esurf}) and (\ref{Evol}). 

\vspace{-2mm}
\section{Condensate Star Limit and the First Law}
\label{Sec:FirstLaw}

In the matching of interior to exterior Schwarzschild solutions the inner and outer surface gravities
$\kappa_{\pm}$ in (\ref{Delkap}) are equal in magnitude, differing only in sign. This is a result of 
the null surface at $r=R_0$ being consistently embedded in a four-geometry, viewed from either
side of the surface interface. Indeed passing to the limit $R\rightarrow R_s^+$ and $R_0 \rightarrow R_s^-$,
and defining the radial coordinate
\vspace{-2mm}
\be
\xi \equiv  \left\{ \begin{array} {ll} -R_s \, \left(1- {\displaystyle \frac{r^2}{R_s^2}}\right)^{\!\frac{1}{2}}\,, 
& \qquad r \le R_s  \vspace{2mm}\\ \ \, 2R_s\, 
{\displaystyle\left(1 - \frac{R_s}{r}\right)^{\!\frac{1}{2}}}\,, &\qquad  r \ge R_s \end{array}\right.
\label{xidef}
\ee
which vanishes at $r=R_s$, the full interior plus exterior Schwarzschild solution in the limiting case for
$R= R_0 = R_s$ can be written in the global Rindler-like form
\be
ds^2 = - \frac{\xi^2}{4 R_s^2} \, dt^2 + q^2(\xi)\, d\xi^2 + r^2(\xi)\, d\Omega^2
\label{txi}
\ee
with
\vspace{-7mm}
\be
q(\xi) = \left\{ \begin{array} {ll}{\displaystyle  \frac{R_s}{r} = 
\left(1 -  \frac{\xi^2}{R_s^2}\right)^{\!\!-\frac{1}{2}}} \,,  & \qquad -R_s < \xi \le 0 \vspace{3mm} \\ 
{\displaystyle\frac{r^2}{R_s^2} = \left(1 - \frac{\xi^2}{4R_s^2}\right)^{\!\!-2}} \,, 
& \qquad \ 0 \le \xi < 2R_s \end{array}\right.
\label{defq}
\ee
continuous at $r=R_s, \xi =0$, and $r(\xi)$ determined in each region by the inverse of (\ref{xidef}). 
Since in this form all metric functions are functions of $\xi^2$, the metric and its first derivative are 
continuous (${\cal C}^1$) across the null surface at $\xi =0$. In coordinates (\ref{txi}) the surface 
gravity $\kappa = \kappa_f$ defined by (\ref{surfgrav}) is
\be
\kappa = \frac{1}{q}\, \frac{d}{d\xi} \left(\frac{|\xi|}{2R_s}\right) = \frac{1}{2qR_s} \, {\rm sgn}\,(\xi)
\label{kapxi}
\ee
which because of the continuity of $q \rightarrow 1$ in (\ref{defq}) as $\xi \rightarrow 0$ is approached 
from either side, results in
\vspace{-3mm}
\be
\kappa_{\pm} = \pm  \frac{1}{2R_s}
\label{kapdSS}
\ee
as in (\ref{kappm}) when $R=R_0=R_s$. The coordinates (\ref{txi})-(\ref{defq}) are 
{\it admissable} in the sense of ref. \cite{Lich}

For this ${\cal C}^1$ matching to the exterior Schwarzschild solution the $\frac{1}{4}$ factor 
in the interior de Sitter region (\ref{fsoln}) is essential, which also determines
the surface gravities $|\kappa_+| = |\kappa_-|$ to being  equal in magnitude. Thus the limiting case 
$R\rightarrow R_s^+$ of the constant density interior Schwarzschild solution provides an explicit 
matching of a (modified) de Sitter interior to the Schwarzschild exterior, compatible with general 
requirements of boundary layers in General Relativity \cite{Lich,BonVic}, evading the longstanding 
presumption that such a matching at their mutual Killing horizons is not possible \cite{PoiIsr}.
The precise formulation of the matching across a null surface according to the appropriate
limit of extrinsic curvature tensors is described in more detail in Appendix \ref{Sec:Junction}.

\begin{figure}[t] 
\begin{center}
\includegraphics[height= 4.6cm, trim=2cm 1cm 2cm .6cm, clip=false]{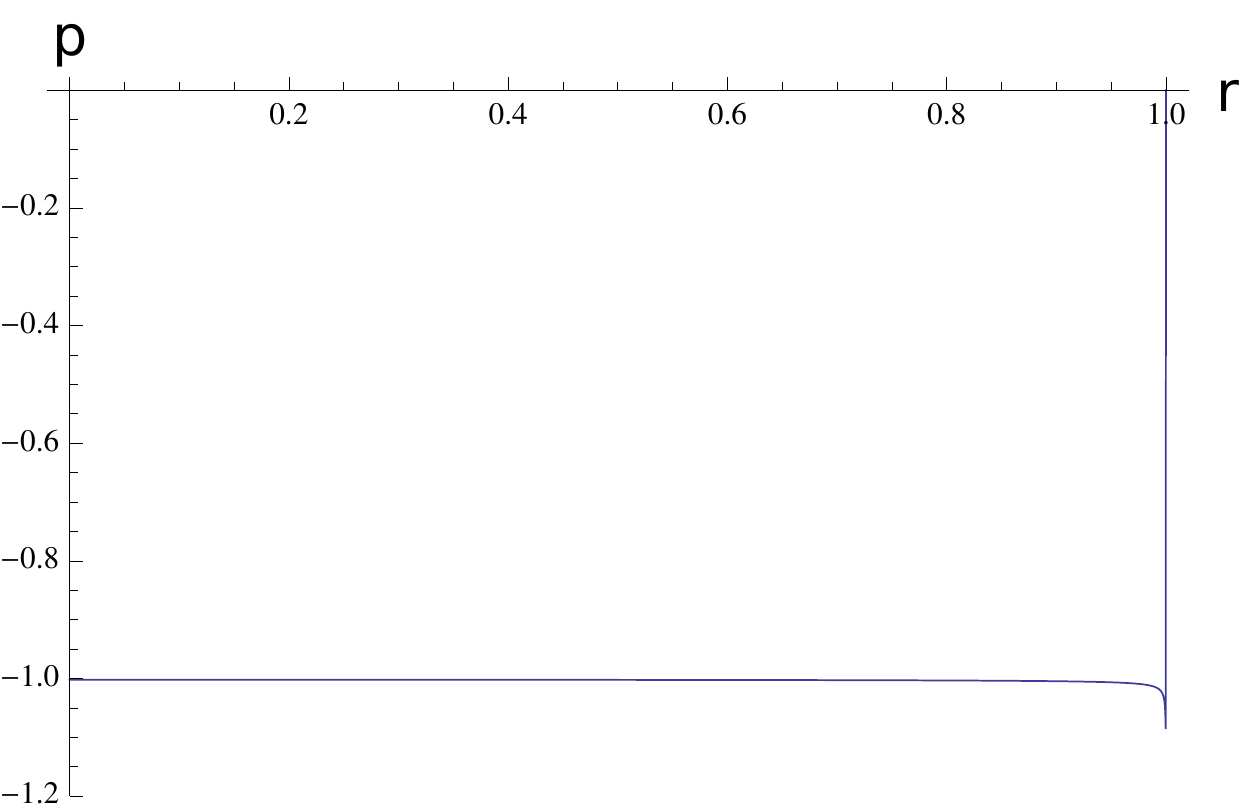}
\end{center}
\vspace{1mm}
\caption{The constant negative pressure $p= -\bar\rho$ as a function of $r/R_s$ of the limiting form
of interior Schwarzschild gravitational condensate star solution for $R= 1.000001\, R_s$.}
\label{Fig:Pressureconst} 
\end{figure}

\begin{figure}[ht] 
\begin{center}
\vspace{4mm}
\includegraphics[height=5cm, trim=2cm 7mm 2cm 1cm, clip=false]{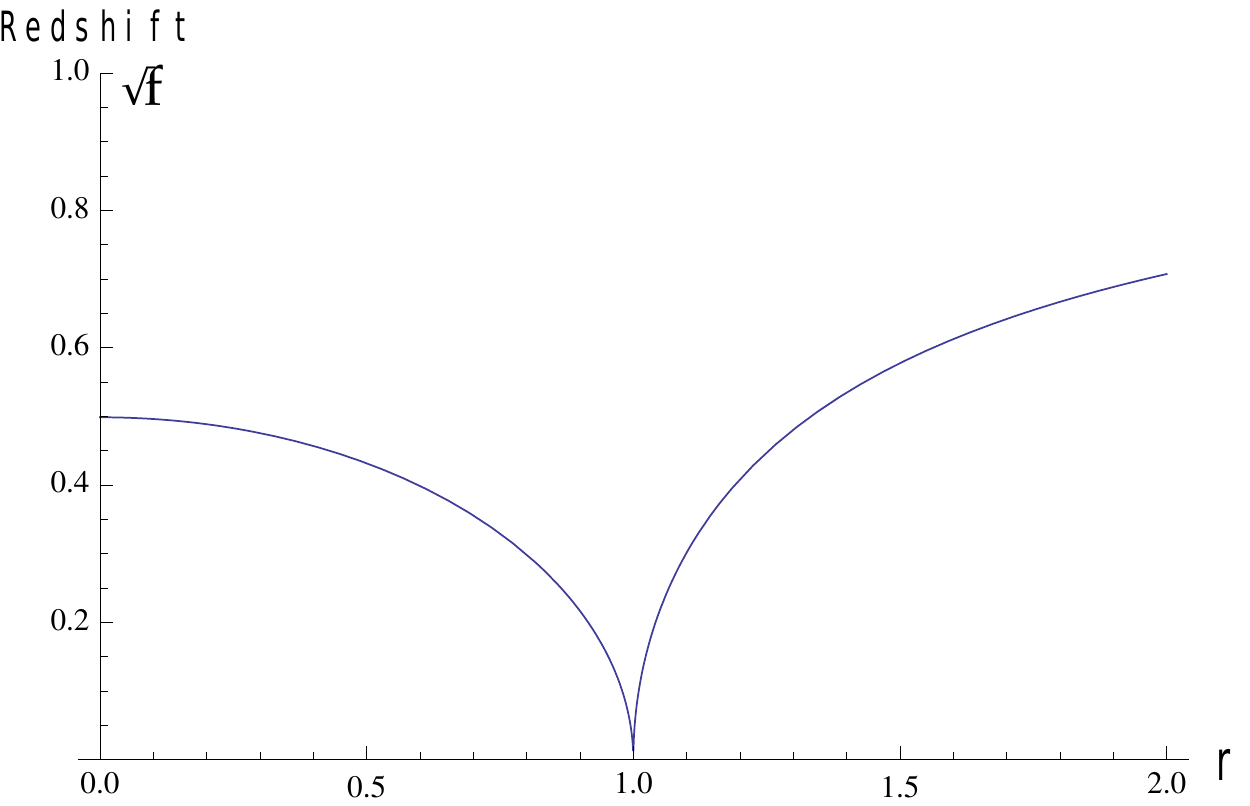}
\end{center}
\caption{The redshift factor $f^{\frac{1}{2}}$ as a function of $r$ (in units of $R$) of the full Schwarzschild
gravitational condensate star solution for $R = R_s$. Note the value of the redshift factor $f^{\frac{1}{2}}(0) = 0.5$
at the center of the star.}
\label{Fig:Redshiftfinal} 
\end{figure}

\begin{figure}[ht] 
\begin{center}
\vspace{-5mm}
\includegraphics[height=5cm, trim=1.6cm 7mm 2cm 1cm, clip=false]{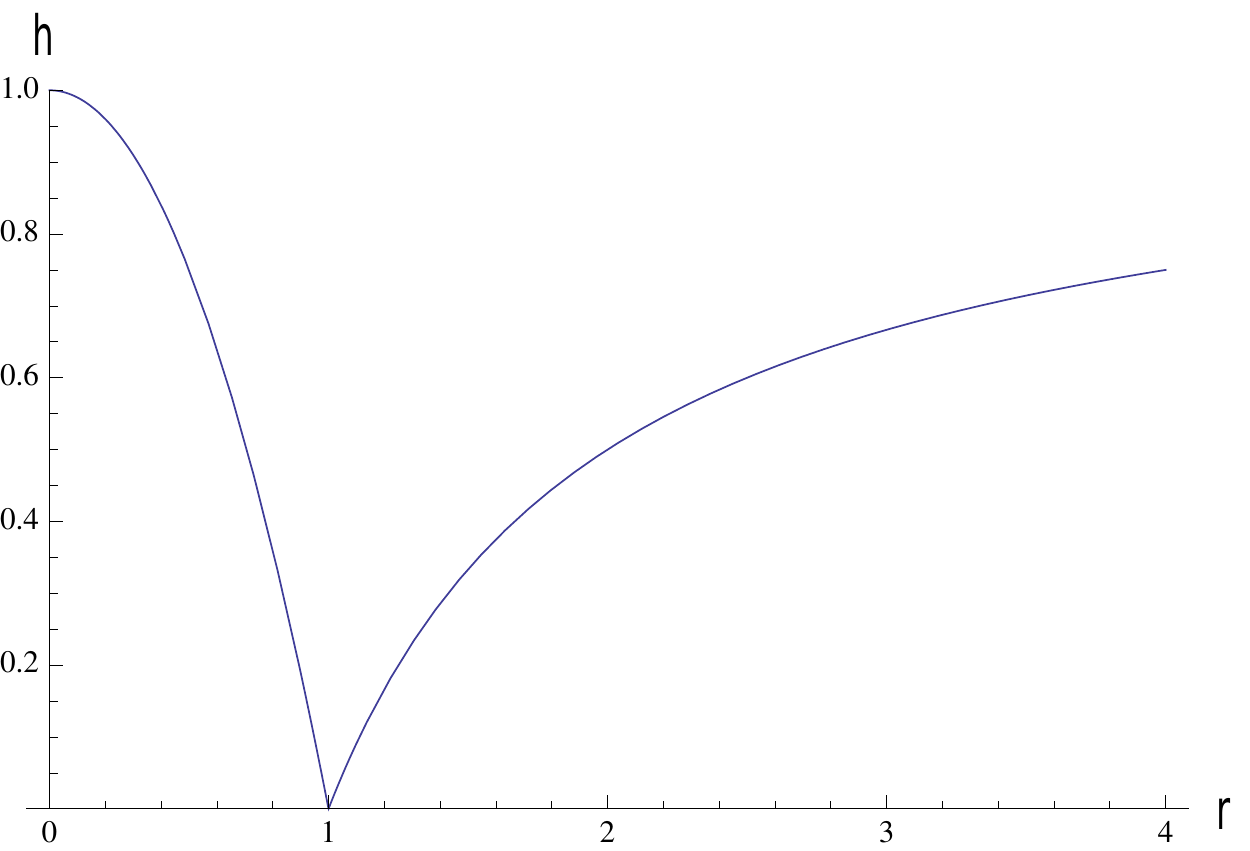}
\end{center}
\caption{The metric function $h$ as a function of $r/R_s$ of the gravitational condensate star solution for $R= R_s$.}
\vspace{-1mm}\label{Fig:hfinal} 
\end{figure}

In the limit $R=R_0=R_s$, (\ref{txi})-(\ref{kapdSS}) apply and the discontinuity of the surface gravities and (\ref{deftau})
give the surface tension of the membrane at the null surface $r=R_s, \xi =0$ 
\be
\tau_s =  \frac{\Delta\kappa}{8 \pi G} \equiv \frac{\kappa_+ - \kappa_-}{8 \pi G} = \frac{1}{8\pi G R_s} = \frac{1}{16 \pi G^2 M}
\label{taus}
\ee
exactly twice the `surface tension' in Smarr's formula for uncharged, non-rotating black holes (\ref{SmarM}). 
That result takes into account only the outer Schwarzschild geometry $\kappa_+$, while $\Delta \kappa = 2 \kappa_+$  
is the actual surface tension of the transvese pressure term $p_{\perp}$ and surface energy (\ref{Esurf}) associated with the 
pressure jump from interior modified de Sitter to exterior Schwarzschild geometries. The formal analogy to a surface 
tension in Smarr's black hole formula (\ref{SmarM}) is thus made precise as a mechanical property of the physical
surface or membrane at $r=R_s$. It is the {\it discontinuity} in surface gravities that gives a true surface energy contribution 
(\ref{Esurf}) to the Komar mass-energy from the $\delta$-function in $p_{\perp}$ (\ref{pdiff}). Note also that $\tau_s$
given by (\ref{taus}) is small in magnitude for large $M$. The behavior of the interior 
Schwarzschild solution in the limiting case with the $p= - \bar\rho$ interior de Sitter solution (modified by the 
$\frac{1}{4}$ factor in $f$) as $R \rightarrow R_s^+$ from above and $R_0 \rightarrow R_s^-$ from below is 
shown in Figs. \ref{Fig:Pressureconst}-\ref{Fig:hfinal}.

In this (gravitational condensate star or gravastar) limit the volume contribution to the Komar mass-energy (\ref{Evol}) 
of the $p = - \bar \rho$ interior is $-M$ and the surface contribution is $2M$ so that
\be
E=M= E_v + E_s = -M + 2 \tau_s A = - M + 2M\,.
\label{intM}
\ee
The surface tension (\ref{deftau}) is then localized in an infinitely thin shell at the Schwarzschild radius itself, 
and the interior solution is a (modified) static patch of de Sitter space within its horizon, as proposed in \cite{PNAS}. 
Eq.  (\ref{intM}) expresses the integral form of the total mass-energy of a gravitational
condensate star as a volume contribution $E_v$ from the $p=-\bar\rho$ interior and a
surface contribution $E_s$ from the thin shell or membrane at $r=R_s$. It therefore takes the place
of the Smarr formula (\ref{SmarM}) for a non-rotating black hole. The differential form of (\ref{intM})
may also be derived by the methods of \cite{BarCarHaw}, by which we find
\vspace{-5mm}
\be
\delta M = \delta E_v + 2\, \tau_s\, \delta A + 
\frac{2}{8 \pi G} \,\big(\delta \kappa_+ - \delta \kappa_-\big)\, A + \left(-\delta M - \frac{A}{4 \pi G}\, \delta \kappa_+\right)
\vspace{-1mm}
\ee
where in the last parentheses the (zero) variation of the Komar energy in the exterior Schwarzschild region 
has been written as a surface integral as in \cite{BarCarHaw}, and (\ref{deftau}) has been used. Thus the
terms involving the variations $ \delta \kappa_+$ cancel. However since
\vspace{-1mm}
\be
- \frac{2}{8 \pi G}\,A\  \delta \kappa_-  = -\frac{4 \pi R_s^2}{4\pi G}\ \delta\!\left(- \frac{1}{2R_s}\right) 
= - \frac{1}{2G} \, \delta R_s = - \delta M = +\delta E_v
\ee
the term with the variation of $\delta \kappa_-$ adds with $\delta E_v$, and we obtain the differential 
mass-energy formula
\vspace{-9mm}
\be
d M =  d E_v +  \tau_s\, d A 
\label{newfirst}
\ee
for the energy conservation (First Law) of a non-rotating gravitational condensate star, in place of (\ref{dM}) 
for a black hole. The form of (\ref{newfirst}) and the previous discussion of the redshifted surface tension 
arising from the transverse pressure $p_{\perp}$ in the Komar formula now fully justifies the identification
of the coefficient of $\delta A$ in (\ref{newfirst}) as the surface tension of the physical surface
located at $r=R_s$ for the full (interior + exterior) Schwarzshild solution in the limit $R\rightarrow R_s$.

It bears emphasizing that the surface energy contribution $E_s$ in (\ref{Esurf}) and (\ref{intM})
does not arise from any energy density $T_{tt}$ at the surface, but purely from the transverse 
anisotropic pressure $T^{\theta}_{\ \theta} = T^{\phi}_{\ \phi} = p_{\perp}$ which contributes to the 
Komar energy (\ref{KomarM}). If the Misner-Sharp energy (\ref{mint}) is used in (\ref{intM}) and (\ref{newfirst}) 
instead, then $\delta M = \bar\rho \ \delta V_s = - p\ \delta V_s$ with $\delta V_s = 4 \pi R_s^{\ 2} \,\delta R_s$, 
which ascribes {\it all} of the energy to the volume contribution of constant negative pressure within $r<R$.
Gravitational energy cannot be uniquely localized. Only the total energy of the system (matter 
plus gravitational) is well defined. Thus whereas the Misner-Sharp mass $m(r)$ and Komar integrand 
are quite different locally, their total integrals give the same Schwarzschild mass and $M =E_v + E_s$ agree, 
as seen from the exterior region. 

We close this section with several additional remarks. 
\begin{itemize}
\vspace{-4mm}
\item Since by the Gibbs relation
\vspace{-6mm}
\be
p + \rho = s\,T + \mu\, n
\label{Gibbs}
\vspace{-4mm}
\ee
and $\mu=0$, no chemical potential corresponding to a conserved quantum number having entered our classical
considerations, the interior Schwarzschild-de Sitter solution with $p + \rho = 0$ is a
zero entropy density $s=0$ and/or zero temperature macroscopic state, justifying its designation 
as a condensate. 
\vspace{-4mm}
\item The differential relation (\ref{newfirst}) expressing the conservation of mass-energy is purely a
{\it mechanical} relation, entirely within the domain of classical General Relativity, rather 
than a quantum or thermodynamic relation. The area $A$ is the geometrical  area of the condensate star 
surface with no implication of entropy. There is no entropy at all associated with a macroscopic condensate 
at zero temperature, as (\ref{Gibbs}) shows. The Planck length $L_{Pl} =\sqrt{\hbar G/c^3}$ or Planck mass 
$M_{Pl} = \sqrt{\hbar c/G}$ or $\hbar$ have not entered our considerations at all up to this point.  
\vspace{-4mm}
\item The matching of the metric interior to the exterior solution for $R=R_s$ has the 
cusp-like behavior shown in Figs.\,\ref{Fig:Redshiftfinal}-\ref{Fig:hfinal}, which is non-analytic in 
the original Schwarzschild radial coordinate $r$, invalidating the assumption of complex metric 
analyticity needed for deriving periodicity in imaginary time $t$. Unlike in the analytically extended 
vacuum Schwarzschild solution, where $f(r)$ becomes negative and the Killing vector $K^{\mu}$ 
becomes spacelike in the interior $r < R_s$ of the black hole, in the negative pressure Schwarzschild-de Sitter 
interior solution with surface tension (\ref{deftau}) at $R=R_s$, complex analytic continuation is not possible, 
and there is no requirement of any fixed periodicity in imaginary time. Thus despite its relation to acceleration, 
the surface gravity $\kappa$ carries no implication for temperature or thermal radiation, and its discontinuity 
is a purely mechanical property, namely the surface tension of a physical boundary layer at $r=R_s$.
\vspace{-4mm}
\item Non-analyticity at $r=R_s$ is exactly the property suggested by the analogy of black hole 
horizons to phase boundaries and quantum critical surfaces in condensed matter physics \cite{CHLS,Lau}. 
Since $f(r) = 0$ corresponds to the `freezing' of local proper time at $r=R_s$, it suggests critical slowing down
characteristic of a phase transition. The vanishing of the effective speed of light $c^2_{eff} =f(r)$ is analogous 
to the behavior of the sound speed determined by the low energy excitations at a critical surface or phase boundary.
This suggests in turn that gravitation and spacetime itself are `emergent' phenomena of a more 
fundamental microscopic many-body theory \cite{Mazur97,Lau}. 
\vspace{-4mm}
\item The positivity of $c^2_{eff} \equiv f(r)$ on either side of the phase boundary and its interpretation 
as the effective speed of light squared, which must always be non-negative, brings to mind Einstein's 
original papers on the local Relativity Principle for static gravitational fields, which led him to General 
Relativity from the Minkowski metric $ds^2 = -c^2\, dt^2 + dx^2 + dy^2 + dz^2$, by allowing first the time 
component $-g_{tt} = c^2$ and eventually all other components of the metric to be functions of space (and 
in general also time) \cite{Ein12}. Thus it could be argued that the non-negativity of $c^2_{eff} =f(r)$ in
a static geometry, and first order differentiability of the metric in the Rindler-like coordinates 
(\ref{txi}), is more faithful to Einstein's original conception of the Equivalence Principle, realized by
real continuous coordinate transformations, than is complex analytic extension around a square
root branch point that would allow $c_{eff}^2 <0$. At the minimum, the matching of the $p=-\bar\rho$ 
Schwarzschild interior to exterior provides a consistent logical alternative to analytic extension, 
entirely within the framework of classical General Relativity, provided only that surface boundary 
layers on null boundaries are admitted.
\vspace{-4mm}
\item Finally,  since $K^{\mu}$ remains timelike for a gravastar, $t$ is a global time and unlike in
the analytic continuation hypothesis, the spacetime (\ref{txi}) is truly static. The $t=$ const. hypersurface 
is a Cauchy surface and is everywhere spacelike. This is exactly the property of a static spacetime necessary 
to apply standard quantum theory, for the quantum vacuum to be defined as the lowest energy
state of a Hamiltonian bounded from below, and for the Schr\"odinger equation to describe 
unitary time evolution, thus avoiding any possibility of an `information paradox.'
\end{itemize}
\vspace{-4mm}

\section{Defocusing of Null Geodesics and Surface Oscillations}
\label{Sec:Observ}

Since the Schwarzschild time $t$ is a global time, and $f(r)$ does not change sign in the
interior Schwarzschild solution with $R=R_0=R_s$, there is no event horizon. The touching
of zero of the cusp at $r=R_s$ in Fig. \ref{Fig:Redshiftfinal} will almost certainly be removed
in a more complete theory, such as suggested by the $\eps$ regulator introduced in
Appendix \ref{Sec:DeltaDistrib}. General semi-classical estimates lead one to expect that
$\eps^2 \sim L_{Pl}^2/R_s^2  \propto \hbar$, in which case $f(r)$ would be very small in
the vicinity of $R_s$ but nonetheless strictly positive everywhere. Even in the limiting 
case of $\eps \rightarrow 0$, where $f(r)$ vanishes at $r= R_s$, light rays with any finite 
positive radial momentum are still able in principle to pass from the interior outward 
through the Schwarzschild sphere to the exterior region. Since this is very different
from the behavior of light rays trapped inevitably by a black `hole,' the possibility
arises of distinguishing a gravastar from a black hole by optical imaging,  {\it e.g.} by
VLBI in the near infrared \cite{EHT,SakSaiTam}.

The behavior of light rays in the full $p= - \bar\rho$ interior plus exterior geometry can
be studied by means of the geodesic equation
\be
-\frac{E^2}{f(r)} + \frac{1}{h(r)} \left(\frac{dr}{d\lambda}\right)^2 + \frac{L^2}{r^2} + m^2 = 0
\label{geodesic}
\ee
for zero mass particles $m^2 = 0$. The constants of the motion $E$ and $L$, energy and
angular momentum respectively, are defined in terms of the canonical momenta $p_{\mu}$ and 
the affine parameter $\lambda$ along the trajectory by
\vspace{-3mm}
\bes
\bea
E &\equiv& -p_t = f(r) \, \frac{dt}{d\lambda}\\
L^2 \equiv p_{\theta}^2 + \frac{p_{\phi}^2}{\sin^2 \theta} &=&
r^4\left[ \left(\frac{d\theta}{d\lambda}\right)^2 + \sin^2\theta\, \left(\frac{d\phi}{d\lambda}\right)^2\right]\,.
\eea
\label{canmom}\ees
Thus the null geodesic eq. may be written in the form
\be
\left(\frac{dr}{d\lambda}\right)^2 + {\cal V}(r) = 0
\label{nullgeod}
\ee
in terms of the effective radial potential
\be
{\cal V}(r) =\frac{h(r)}{r^2}\, L^2- \frac{h(r)}{f(r)}\, E^2
\label{effV}
\ee
which shows that only light rays with vanishing energy $E =0$ can hover indefinitely at $r=R_s$.

Since angular momentum is conserved, the motion takes place in a plane, which without loss
of generality can be chosen to be the equatorial plane at $\theta = \frac{\pi}{2}$. Then 
$p_{\phi} = L = r^2 \frac{d\phi}{d\lambda}$, and by dividing (\ref{nullgeod}) 
by $L^2$ for non-radial geodesics, and defining in the usual way the variable
\be
u \equiv \frac{R_s}{r}\qquad {\rm and} \qquad b \equiv \frac{L}{E}
\label{defu}
\ee
the impact parameter, the equation for the null trajectory of a photon may be written as\\

\vspace{-1.5cm}
\be
\left(\frac{du}{d \phi}\right)^2 = \bigg[\frac{R_s^2}{b^2} \,\frac{h}{f} - u^2\, h\bigg]_{\displaystyle r = \tfrac{R_s}{u}}
\label{nulltraj}
\ee
in plane polar coordinates. Substituting the interior solution (\ref{fhint}) for $R=R_0=R_s$ we obtain
\be
\left(\frac{du}{d \phi}\right)^2 = \frac{4R_s^2}{b^2} + 1 - u^2\,,\qquad u> 1
\label{nullint}
\ee
whose general solution with $\displaystyle\frac{du}{d\phi} \neq 0$ is
\be
u = \sqrt{1 + \frac{4R_s^2}{b^2}}\, \cos(\phi - \phi_0)\,,\qquad r< R_s\,,\qquad 1 \le u \le \sqrt{1 + \frac{4R_s^2}{b^2}}
\label{uintsoln}
\ee
in the (modified) de Sitter interior. The upper bound on $u$ corresponds to the radius 
\be
r_{min} = \ \frac{\ b\, R_s}{\!\!\sqrt{b^2 + 4R_s^2}} \,\le \,R_s
\ee
of closest approach to the origin in the interior, which is achieved at $\phi= \phi_0$. At angles
\be
\phi_{\pm} = \phi_0 \pm \sin^{-1}\left(\,\frac{2R_s}{\!\!\sqrt{b^2 + 4R_s^2}}\right)
\ee
the null ray enters and exits the interior region.

Likewise in the exterior Schwarzschild region one obtains from (\ref{nulltraj}) the photon trajectory
\be
\left(\frac{du}{d \phi}\right)^2 = \frac{R_s^2}{b^2} - u^2 + u^3 \,,\qquad r>R_s\,,\qquad 0<u\le 1
\label{nullext}
\ee
which is solved by the Weierstrass elliptic function ${\cal P}(\phi; g_2, g_3)$ in the form \cite{Mazur97}
\be
u = \tfrac{1}{3} + 4 \, {\cal P}(\phi; g_2, g_3)
\label{uextsoln}
\ee
in terms of the elliptic invariant parameters
\be
g_2 = \frac{1}{12}\,,\qquad g_3 = \frac{1}{216} - \frac{R_s^2}{16\, b^2}
\label{g1g2def}
\ee
and we have set a possible second integration constant to zero by choice of the axis at
which $\phi = 0$.

\begin{SCfigure}[0.6][ht]
\includegraphics[height=0.3\textheight, width=0.4\textwidth]{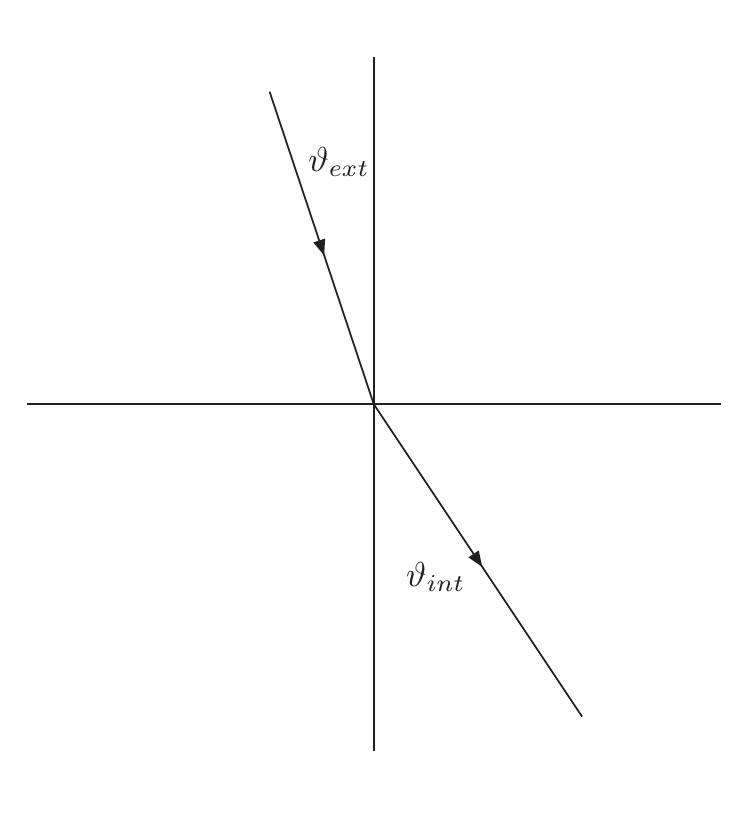}
\hspace{1cm}
\caption{The refraction of a null ray at the surface of a gravitational condensate star.
The angles $\vartheta_{ext}$ and $\vartheta_{int}$ are the angles the null ray makes
with the normal to the surface in the Schwarzschild exterior and (modified) de Sitter interior
respectively.  Since according to (\ref{refract}) $\vartheta_{ext} < \vartheta_{int}$, the 
$p= - \bar\rho$ condensate interior behaves as a medium with an index of refraction less than unity.}
\label{Fig:Refract} 
\end{SCfigure}

The important feature of these null geodesics is that they can penetrate and re-emerge from
the surface at $r=R_s$. By evaluating (\ref{nullint}) and (\ref{nullext}) at the boundary $u=1$ we
see that $\displaystyle \tan \vartheta = \Big\vert\frac {du}{d\phi}\Big\vert$ so that they are refracted 
there according to Snell's Law
\be
\sin \vartheta_{ext}\ \sqrt{1 + \frac{4R_s^2}{b^2}}\  =\  \sin\vartheta_{int}\ \sqrt{1 + \frac{R_s^2}{b^2}}
\label{refract}
\ee
with $\vartheta_{ext}$ and $\vartheta_{int}$ the angle the light ray makes with the normal to the
surface at $r=R_s$ in the exterior and interior respectively. This is illustrated in Fig. \ref{Fig:Refract}.

\begin{figure}[ht]
\begin{center}
\vspace{4mm}
\includegraphics[height=4.5cm, trim=3cm 3cm 3cm 2cm, clip=false]{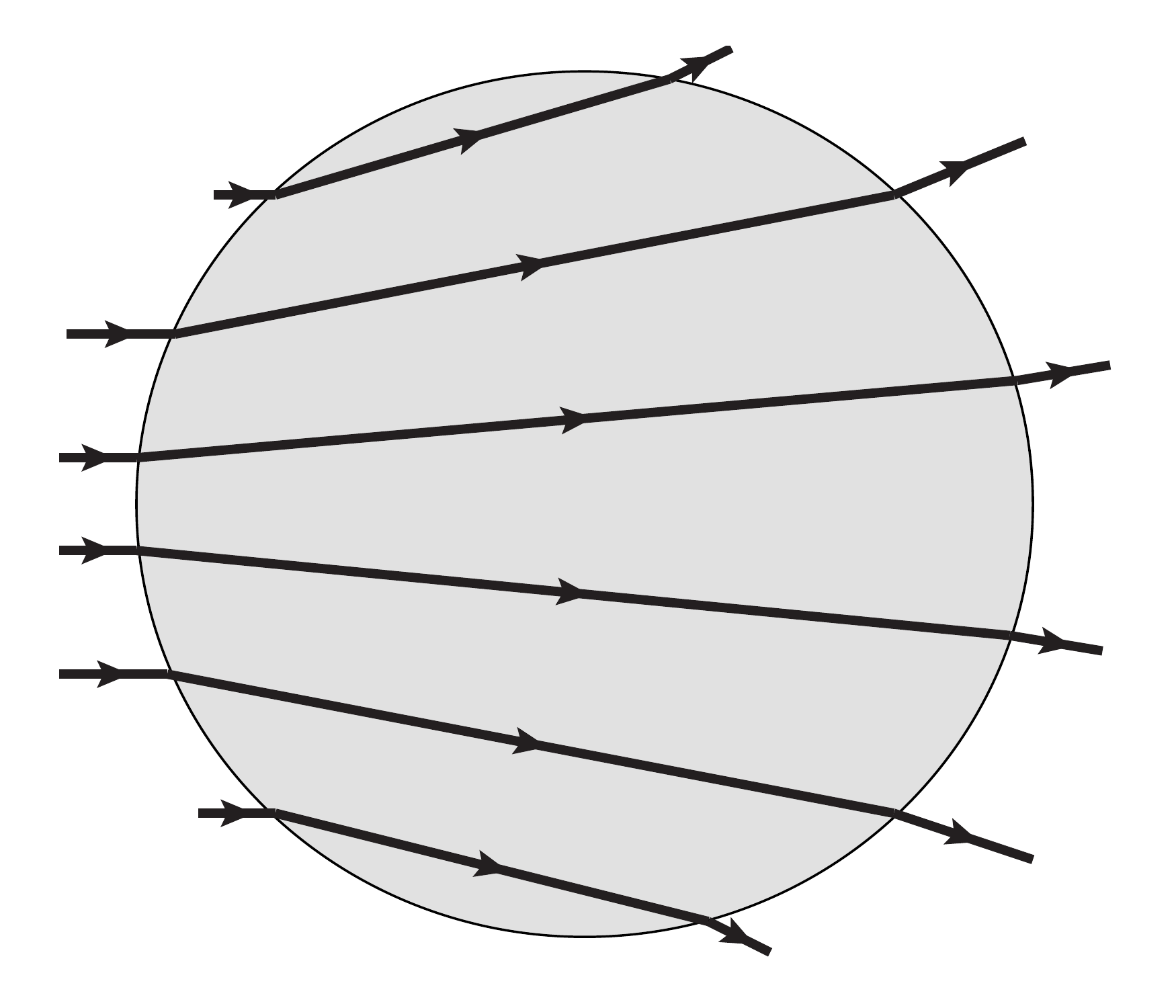}
\end{center}
\hspace{1cm}
\caption{The defocusing of null rays passing through the interior of a gravitational condensate star.}
\label{Fig:Defocus} 
\end{figure}

Since $\vartheta_{ext} \le \vartheta_{int}$ with equality attained only for radial geodesics with $b=0$
and $\vartheta_{ext}= \vartheta_{int} = 0$, the condensate interior acts as medium with index 
of refraction $n <1$, or {\it negative} lens with respect to the vacuum exterior. Hence light rays are {\it defocused} 
by passing through the interior as illustrated in Fig. \ref{Fig:Defocus}, and a gravitational condensate star will have
optical imaging characteristics quite distinct from a black hole which absorbs all light impinging on its horizon. 
The detailed imaging expected clearly merits a full analysis and modeling in realistic astrophysical 
environments for comparison to observations.

A second important qualitative difference between gravitational condensate stars and black holes is
the existence of surface dynamics. In the strictly classical approximation of the interior Schwarzschild
solution, the surface has zero thickness and additional information about the composition of the surface
is necessary to determine the normal modes of oscillation. This requires ideally a Lagrangian model 
for the degrees of freedom of the surface, or at the least a phenomenological parameterization of the 
restoring and damping forces acting upon it. A phenomenological treatment of gravitational
waves due to inspiral was given in ref. \cite{PBCCN}. A complete analysis of stability
of gravitational condensate stars awaits a full dynamical theory. A necessary building
block of that more complete theory was proposed in \cite{MotVau,Zak}, namely the effective action
of the quantum conformal anomaly, which possesses an additional scalar degree of freedom coupling
strongly at the horizon or gravastar surface.

Even in the absence of a complete theory, it is clear on physical and dimensional grounds that the natural 
frequency of oscillation of the surface modes must be set by the size of the condensate star $R_s$, so that
\be
\omega \sim \frac{c}{R_s} = 101.5 \ \left(\frac{M_{\odot}}{M}\right) \, {\rm kHz}
\label{oscfreq}
\ee
is the relevant frequency scale. Because the surface is closed, and the surface oscillations will generally 
have a non-zero quadrupole moment, they will generate gravitational waves at characteristic {\it discrete} 
quasi-normal mode frequencies of the order of (\ref{oscfreq}). This is quite a distinct gravitational wave signature 
from the the infalling, coalescence and chirp modes expected for black holes without a physical surface  \cite{ChiRez}. 
Since the optimal frequency sensitivity of the Advanced LIGO-LSC gravitational wave interferometers lie between 
approximately $50$ Hz and $750$ Hz,  this corresponds to black hole/gravastar masses in the range of 
$10^2\, M_{\odot}$ to a few times $10^3\, M_{\odot}$ \cite{LIGO}. 

Further observational tests for distinguishing gravastars with a surface from black holes are possible
in the case case of rotating gravastars, which have been considered in \cite{HarKovLob}, and will be taken
up in more detail future work.

\section{Conclusions and Outlook}
\label{Sec:Conclusions}
\vspace{-1mm}

The constant density interior Schwarzschild solution possesses some remarkable properties that
seem not to have been appreciated for nearly a century. Its importance lies not in that any ordinary 
matter can be incompressible and exist at constant density under the extreme pressures
of gravitational collapse, but in the fact that it provides a simple extreme state that saturates
the bounds applicable to much more general, spherically symmetric configurations; and
in the interesting physical insights and limiting behaviors it provides. In particular, the Buchdahl 
bound $R > \frac{9}{8} R_s$ is based on a comparison with the constant density 
interior Schwarzschild solution. Hence the behavior of the central pressure in this solution 
is expected to be applicable to other equations of state, and to diverge in a similar 
way when a spherically symmetric star with a more realistic eq. of state first contracts to
an even larger radius outside of its Schwarzschild radius. 

By study of the constant density interior Schwarzschild solution we have shown that negative
pressure is produced in the center of a spherically symmetric star well before its Schwarzschild 
radius is reached or a trapped surface is formed. Rather than being a reason to reject the 
solution for $R \le \frac{9}{8} R_s$, we have shown that the pressure divergence is integrable, 
according to the Komar mass-energy (\ref{KomarM}) for a static, spherically symmetric star. A
surface energy density localized at the radius (\ref{R0}) is necessarily produced, with a 
finite redshifted surface tension given by (\ref{taus}) in terms of the discontinuity of the surface 
gravities. The crucial observation is that the static Killing vector $K^{\mu}$ becomes null and 
$-K^{\mu}K_{\mu} = f(r)$ vanishes at exactly the same radius as the pressure divergence.
Since $\sqrt{f\,}$ multiplies the pressure in the Komar energy integral (\ref{KomarM}), the result 
is that an integrable $\delta$-function of transverse stress is generated at that radius. With this 
modification, the isotropy assumption $p_{\perp} = p$ upon which the Buchdahl bound is based 
is evaded, and the interior Schwarzschild solution becomes again a viable model for the non-singular
interior of a fully collapsed star, particularly in the limit $R\rightarrow R_s$.

As $R \rightarrow R_s^+$ from above, the surface discontinuity moves out to the Schwarzschild 
radius itself, $R_0 \rightarrow R_s^-$, and most remarkably the interior becomes one of uniformly 
constant negative pressure $p= - \bar\rho$. Thus the vacuum dark energy eq. of state emerges 
naturally from classical  General Relativity under enough spherical compression of matter, 
and prior to the formation of any trapped surface or event horizon. This suggests that although
the sequence of constant density configurations for $R \le \frac{9}{8} R_s$ may not be physically 
realistic in detail, the limiting case of $R=R_0=R_s$ may have much broader applicability, 
indicating that a divergence in the central pressure or curvature singularity is avoided by a 
phase transition to a negative pressure condensate and formation of a phase boundary 
between the positive and negative pressure regions. Since in the adiabatic limit, spherical 
gravitational collapse may be conceived as passing through a sequence of slowly decreasing 
equilibrium states of fixed $R$, it further suggests that the phase transition to $p= - \bar\rho$ 
occurs first at the center of the star and moves {\it inside out}, resulting in an explosive event that would 
expel prodigious amounts of energy and entropy. The final quiescent state of complete 
gravitational collapse to $R_s$ may be a gravitational condensate star with a $p= - \bar\rho$ 
modified de Sitter interior, with a finite surface tension, rather than a black hole. 

Although a fully satisfactory description no doubt requires quantum theory, 
it is remarkable that the possibility, or the {\it prediction}, of a phase transition to a negative pressure 
$p= - \bar\rho$ equation of state counteracting the attractive force of gravity and preventing a singularity 
in gravitational collapse exists already in Einstein's classical theory, independently of the detailed 
composition of the matter being compressed. This may be less surprising if General Relativity
is an effective low energy theory which is a limiting case or `emergent'  from a more fundamental 
microscopic description that takes full account of the quantum nature of both matter and 
spacetime \cite{Mazur97}.
\vspace{-1mm}

From the perspective of the general theory of boundary layers and junction conditions in 
General Relativity, the interior Schwarzschild solution provides an interesting, explicit example 
showing that a second possibility for joining geometries at a null surface, distinct from the analytic 
continuation assumption through a mathematical event horizon usually adopted, is both
logically and physically possible. Inspection of the line element of the global Schwarzschild 
solution in Rindler-like coordinates (\ref{txi})-(\ref{defq}) shows that the spacetime appears to be
locally flat (except for the $\delta$-fn.) and ${\cal C}^1$ differentiable in the vicinity of the null surface 
at $\xi=0$. The Killing norm $\displaystyle \frac{\xi^2}{4R_s^2}\ge 0$ does {\it not} change sign from 
positive to negative as one passes through $\xi=0$, and this has the consequence that there is a 
$\delta$-function in transverse stress and curvature localized at $\xi =0$. This is very different from 
the analytic continuation hypothesis through complexified $(t,r)$ coordinates, in that Figs. 
\ref{Fig:Redshiftfinal}-\ref{Fig:hfinal} show that the metric functions have a non-analytic
cusp-like behavior at $r=R_s$. 
\vspace{-1mm}

The equality of surface gravity magnitudes $|\kappa_+| = |\kappa_-|$ (\ref {kapdSS}) realized by the explicit interior 
Schwarzschild example shows that the matching of a static de Sitter interior (modified by the $\frac{1}{4}$ in
$g_{tt}$) to the Schwarzschild vacuum exterior is possible, evading the long presumption that such a 
matching would necessarily produce an unacceptable metric or curvature singularity \cite{PoiIsr}. 
Instead there is a $\delta$-function distributional transverse pressure in an infinitely thin surface layer, 
leading to a finite surface tension and integrable finite surface energy, after accounting for the kinematical 
redshift factor $\sqrt{-g_{tt}}= \sqrt{f\,}$. The non-analytic behavior of the redshift factor, touching zero (classically) 
at  $r=R_0$, but otherwise everywhere positive, and the sudden change in the vacuum energy from zero to
$\bar\rho >0$ across the boundary at $r=R_s$ are both strongly suggestive of a quantum phase transition and 
the behavior expected across a quantum critical surface, again pointing towards a more fundamental quantum 
many-body theory of gravitation \cite{Mazur97}, and in analogy with similar behaviors in more familiar examples 
in condensed matter systems \cite{Lau}. The eq. of state of constant $p= -\bar\rho$ is exactly that which should 
be expected for a quantum macroscopic state with zero entropy and temperature from the Gibbs relation
(\ref{Gibbs}) with zero chemical potential, and hence of a coherent gravitational Bose-Einstein condensate 
of an underlying many-body theory, described here in classical terms.

In \cite{PNAS} the factor multiplying the interior de Sitter time in static coordinates was treated 
as an unknown constant $C$ to be determined by the matching through a finite boundary layer. 
The present treatment shows that in order to match properly to the interior de Sitter region, $C= \frac{1}{4}$ 
is required in the limit of an infinitely thin boundary layer at $r=R_s$. This results in the interior de Sitter 
time at $r=0$ running at {\it half the rate} of the asymptotic Schwarzschild time $t$ at infinity.

The discontinuity of the surface gravities $\kappa_+ - \kappa_- = 2 \kappa_+$ across the boundary
surface also allows for a completely mechanical and classical interpretation of the mass-energy 
changes of a gravastar expressed in the First Law (\ref{newfirst}). The surface area $A$ in this 
relation does not acquire any interpretation in terms of entropy, and indeed no such interpretation 
is possible for a condensate interior which has zero entropy. Likewise surface gravity 
acquires no interpretation in terms of temperature, and indeed no such interpretation is
possible given the non-analytic behavior of the metric functions at the cusp. The final 
macroscopic quantum state attained by a spherical body collapsed to its Schwarzschild
radius, like that of a neutron star, is one of absolute zero temperature. 
\vspace{-1mm}

Since the exterior and interior metrics match through (\ref{txi})-(\ref{defq}), the Schwarzschild real 
Killing time coordinate $t$ is a global time, and there is no requirement of periodicity in imaginary 
time which would lead to a thermal field theory interpretation, or black hole radiance. The proper 
interpretation of the discontinuity of surface gravities as surface tension in the extended Smarr 
formula (\ref{SmarM}) and First Law (\ref{dM}), previously lacking, requires a well-behaved 
interior solution, and the correct matching of the Schwarzschild exterior solution to the (modified) 
de Sitter interior across a null surface, as described in Appendix \ref{Sec:Junction}.
Since the $t=$ constant hypersurface is a Cauchy surface and everywhere spacelike,
standard quantum theory can be applied, the quantum vacuum is well defined as the lowest energy
state of a Hamiltonian bounded from below, and the Schr\"odinger equation describes 
unitary time evolution, thus avoiding any possibility of an `information paradox.'

In this paper we have made no attempt to provide a full theory of the phase transition to
the $p= - \bar\rho$ condensate or the surface boundary layer. Instead our aim has been 
to show how far one can go to describe a gravitational condensate, negative pressure 
interior and surface tension of a fully collapsed state completely within Einstein's classical 
theory, and therefore consistently with the Equivalence Principle, with $c^2_{eff} = f(r) \ge 0$
as Einstein himself first conceived it \cite{Ein12}. 
\vspace{-1mm}

A more complete quantum treatment will contain as one important element the scalar degree(s) 
of freedom identified in the effective theory of the quantum conformal anomaly \cite{MotVau,Zak}, 
which has been shown to be macroscopically relevant, and allow the possibility for the 
vacuum energy (usually called the cosmological `constant') to change \cite{AntMot,MotVau,DE}.
This will almost certainly lead to the infinitesimally thin membrane of the classical
Schwarzschild solution being replaced by a finite, but still very thin surface layer, so that $f(r) > 0$
strictly. Elementary semi-classical estimates for the physical thickness $\ell$ of that quantum 
surface layer indicate that $\ell \sim \sqrt{L_{Pl} R_s}$ \cite{PNAS,Zak}. How this physical regulator,
dependent upon $\hbar$, might enter is already suggested by the mathematical procedure
of regulating the $\delta$-fn. distribution in the transverse pressure, discussed in Appendix
\ref{Sec:DeltaDistrib} with $\eps \sim L_{Pl}/R_s$.  

Lastly, although difficult to distinguish observationally from black holes, gravitational condensate
stars do offer some promising possibilities for future astrophysical tests. We have suggested
that perhaps one of the cleanest tests follows from the defocusing characteristics of the a 
$p= - \bar\rho$ interior, which in principle can be penetrated by light rays, {\it cf.} Fig. \ref{Fig:Defocus}. 
If instead of propagating by geometric optics through the phase boundary at $r=R_s$ light
is scattered in a frequency dependent way, this will also produce imaging and lensing 
characteristics quite different from a black hole. Secondly, the existence of a physical surface 
implies the existence of surface normal modes of oscillation, which as a {\it discrete} spectrum 
on the scale of the characteristic frequency (\ref{oscfreq}) should be distinguishable from the 
ringdown quasi-normal modes and chirp signals computed assuming there is no surface but 
a mathematical event horizon instead. Clearly the quantitative details of these predictions 
require a fuller theory and more complete treatment, which we defer to a future publication.

Only when the dynamics of the surface is fully specified can the stability and normal mode 
spectrum of surface excitations be computed reliably. Likewise, only when the interactions 
of the surface layer with ordinary Standard Model matter is fully specified can the question 
of whether the surface heats up or simply absorbs the accreting matter and incorporates 
it into the condensate interior be addressed. The coupling of surface to volume modes and 
the time scale for the damping of surface oscillations are the important quantities
needed to obtain quantitative answers to these questions.\\

\appendix

\section{The Buchdahl and Related Bounds}
\label{Sec:Buchdahl}

A non-trivial bound on the ratio $GM/R$ follows from the classical Einstein eqs. alone,
and was found by Buchdahl \cite{Buch}, with only the apparently mild assumptions of:
\begin{enumerate}[label= (\roman*)]
\item Isotropic pressure: \quad $p_{\perp} = p$
\vspace{-3mm}
\item Monotonically non-increasing density profile: \quad $\displaystyle  \frac{d\rho}{dr} \le 0$
\vspace{-3mm}
\item Continuity of the metric coefficient $f(r)$ and its first derivative $\displaystyle \frac{df}{dr}$ at $r=R$\,.
\end{enumerate}
Under these assumptions, Buchdahl's theorem is that \cite{Buch,Wein,Wald}
\be
R > \frac{9}{8} R_s = \frac{9}{4}\,GM
\label{Bbound}
\ee
is the condition for an interior pressure $p(r) < \infty$ which is everywhere finite and $f(r) > 0$ non-vanishing in the interior 
$r \in [0,R]$. Conversely, if the three assumptions (i)-(iii) hold, and the star is compressed to a radius $R \le \frac{9}{8} R_s$,
then the isotropic pressure $p$ must diverge somewhere in the star's interior.
In fact, as $R \rightarrow \frac{9}{8} R_s$, this divergence appears first at the star's center, since under the same three
assumptions (i)-(iii) one can also prove that the central pressure of the star is bounded from below \cite{MarVis}
\be
p(0) \ge \bar\rho  \left[ \frac{ 1 - \sqrt{1 - R_s/R}} {3 \sqrt{1 - R_s/R} - 1} \right]
\label{p0}
\ee
and the lower bound diverges when $R = \frac{9}{8} R_s$. Moreover the same three assumptions (i)-(iii) are also
sufficient conditions to prove that the metric coefficient $f(r)$ is bounded from above by  \cite{MarVis}
\be
f(r) \le  \frac{1}{4} \left[3  \sqrt{1 - H^2 R^2} - \sqrt{1 - H^2 r^2}\,\right]^2
\label{f0}
\ee
in terms of the mean density (\ref{constden}) with $H^2$ defined in (\ref{defH2}). These results show that as 
$R \rightarrow \frac{9}{8} R_s$ both the divergence of the central pressure $p(0)$ and the freezing
of time $f(0)= 0$ first occur together quite generally at the center of the star, $r=0$.  

The inequalities are obtained from the second condition (ii)  $\rho' \le 0$ above, and are saturated by the constant
density profile $\rho' = 0,\  \rho(r) = \bar\rho$, eq. (\ref{constden}) for a star of total mass $M$ and radius $R$. 
Because of these inequalities established for the general spherically symmetric static solution obeying conditions 
(i)-(iii), compared to the constant density Schwarzschild interior solution, the behavior of that solution as the Buchdahl 
limit $R \rightarrow \frac{9}{8} R_s$ is approached is of fundamental interest.

The logical implication of the Buchdahl theorem is that either:
\vspace{-1mm}
\begin{itemize}
\item The central pressure diverges and a true curvature singularity (black hole) is produced; or
\vspace{-2mm}
\item One of more of the assumptions (i)-(iii) of the theorem must be relaxed.
\end{itemize}
\vspace{-1mm}
In particular, the way the Buchdahl bound (\ref{Bbound}) is evaded by the constant density interior
solution through the relaxation of condition (i), first at the origin, is instructive and may be expected to be quite 
generic. The study of the behavior of the interior Schwarzschild solution (\ref{psoln})-(\ref{fsoln}) for 
$R\rightarrow R_s$ points to the physical resolution, and explicit realization of the second possibility 
of a non-singular interior with an integrable surface energy and surface tension, wherein condition 
(i) is relaxed.
\vspace{-1mm}

\section{Regulating the Singular Distribution at $r=R_0$}
\label{Sec:DeltaDistrib}

The naive evaluation of the right side of (\ref{locr}) with the singular interior solution (\ref{psoln}),
(\ref{fsoln}) gives (\ref{rhslocr}), at least for $r \neq R_0$. The evaluation of the
left side (\ref{lhslocr}) contains an additional local $\delta$-function contribution.
In the Appendix we show that if the singular distributional solution is regulated properly,
it also gives a $\delta$-function contribution in the limit that the regulator is removed,
in agreement with (\ref{lhslocr}). We treat the case $R_0 < R$ first and take the limit 
$R \rightarrow R_s^+, R_0 \rightarrow R_s^-$ at the end.

Let us define the dimensionless radial distance $x$ from the singularity of (\ref{psoln}) by
\vspace{-1mm}
\be
r = R_0\,(1 + x)\,.
\vspace{-1mm}
\label{rxdef}
\ee
We observe that for $R_0 < R$ the singularity in $p(r)$ is a linear divergence in $x \rightarrow 0$ since 
the denominator (\ref{denom}) can be expanded in Taylor series 
\vspace{-5mm}
\be
D(r) = c_1 x\,(1 + c_2\, x + \dots)
\label{Dexpand}
\ee
with
\vspace{-4mm}
\be
c_1 \equiv R_0 \frac{dD}{dr}\bigg\vert_{R_0} = \frac{\ H^2R_0^{\ 2}}{\!\sqrt{1-H^2R_0^{\ 2}}}\,,\qquad c_2 = \frac{1}{2 (1 -H^2R_0^{\ 2})}\,,
\qquad {\rm etc.}
\ee
Since the leading term in $D^{-1}$ is a linear divergence at $x =0$, which defines a distribution 
odd under reflection $x \rightarrow - x$ about the singular point, we regularize this
divergence by the replacement
\be
p + \bar \rho = \frac{2 \bar \rho \sqrt{1-H^2R^2}}{D} = \frac{2 \bar \rho \sqrt{1-H^2R^2}}{c_1\, x}\,\Big( 1- c_2\, x + \dots\Big)
\rightarrow \frac{2 \bar \rho \sqrt{1-H^2R^2}}{c_1}\, \frac{x}{x^2 + \eps^2}\Big(1- c_2\, x + \dots\Big)\,
\label{preg}
\ee
which is the definition of the odd, real Principal Part distribution, $\displaystyle {\cal P}\left(\frac{1}{x}\right)$, that is itself integrable.

Likewise for the function $\Phi (r)$ whose leading singular behavior is even upon reflection  $x \rightarrow - x$ 
about the singular point, we make a similar regulated replacement
\bea
&&\Phi =\frac{1}{2} \ln \left(\frac{D^2}{4\,}\right) \rightarrow \frac{1}{2}\ln \left\{ \frac{c_1^2}{4} \, \big(x^2 + \eps^2\big) 
\big(1 + 2 c_2\, x +\dots\big)\right\}
= \frac{1}{2} \ln \,\left(x^2 + \eps^2\right) + \ln \left( \frac{c_1}{2}\right) +  c_2\, x + \dots\qquad \\
&&\hspace{4cm} f^{\frac{1}{2}} = e^{\Phi} \rightarrow \frac{c_1}{2} \,\big(x^2 + \eps^2\big)^{\frac{1}{2}}\, \big( 1 +   c_2\, x + \dots\big)\,.
\label{fexpand}
\eea
\vspace{-1mm}
Computing now the derivatives we find
\begin{align}
\frac{dp}{dr} &= \frac{ 2\,\bar \rho\, \sqrt{1-H^2R^2}}{R_0\,c_1}\,\frac{d}{dx} \left\{\frac{x}{x^2 + \eps^2}\,
\big( 1 - c_2\,x + \dots\big)\right\}\nn
\vspace{2mm}
&= \frac{2\, \bar\rho \sqrt{1-H^2R^2}}{R_0\,c_1}\,\left[\frac{1}{x^2 + \eps^2} - \frac{2x^2}{(x^2 + \eps^2)^2} 
- \frac{2\,c_2\, \eps^2\, x}{(x^2 + \eps^2)^2} + \dots\right]
\vspace{-5mm}
\end{align}
and\\
\vspace{-1cm}
\begin{align}
(\bar\rho + p)\, \frac{d\Phi}{dr} &=  \frac{2\,\bar\rho \sqrt{1-H^2R^2}}{R_0\,c_1}\,\big(1- c_2\, x + \dots\big)\left(\frac{x}{x^2 + \epsilon^2}\right)
\frac{d}{dx} \left\{ \frac{1}{2} \ln \,(x^2 + \epsilon^2) + \ln \left( \frac{c_1}{2}\right) +  c_2\, x + \dots \right\}\nn
&\hspace{1.5cm}= \frac{2\,\bar\rho \sqrt{1-H^2R^2}}{R_0\,c_1}\,\left[ \frac{x^2}{(x^2 + \epsilon^2)^2} + \frac{c_2\,\eps^2\,x}{(x^2 + \epsilon^2)^2}  + \dots \right]
\end{align}
in the vicinity of the singular point $r=R_0$. Summing these contributions and using (\ref{pperp}) gives
\be
\frac{2\,(p_{\perp} - p)}{r} =  \frac{ 2\bar\rho \sqrt{1-H^2R^2}}{R_0\,c_1}\, \frac{\eps^2}{(x^2 + \eps^2)^2} \, \big( 1 -c_2 \, x + \dots\big)
\label{pperpreg}
\ee
so that the leading potentially singular behavior in $p_{\perp} - p$ vanishes in the limit the regulator $\eps \rightarrow 0$.
This explains why it is not seen if the unregulated expressions for $p$ and $\Phi$ with $\eps =0$ are used uncritically 
as in (\ref{rhslocr}). In addition, since the unregulated functions satisfy $p_{\perp} = p$, the first subleading term proportional to 
$c_2$ in (\ref{pperpreg}) and all subsequent finite terms vanish in the limit $\eps \rightarrow 0$ as well. 

Now the combination that appears in the local form of the mass formula (\ref{locr}) is 
\vspace{-1mm}
\bea
&&8\pi r^2\, \left(\frac{f}{h}\right)^{\frac{1}{2}}\, \big(p_{\perp} - p\big) =  
4\pi r^3 \left(\frac{f}{h}\right)^{\frac{1}{2}}\, \left[ \frac{2\,(p_{\perp} - p)}{r}\right]\nn
&& = 4 \pi R_0^{\ 2}\, \bar\rho \, \frac{\big[1-H^2R^2\big]^{\frac{1}{2}}\, (1 + x)^3}{\big[1-H^2R_0^{\ 2}\,(1 + x)^2\big]^{\frac{1}{2}}}\,
 \,\frac{\eps^2}{(x^2 + \epsilon^2)^{\frac{3}{2}}} \,\big(1+ c_2\,x + \dots +\big)\,\big(1- c_2\, x + \dots \big)\nn
&&= 4 \pi R_0^{\ 2}\, \bar \rho \, \frac{\sqrt{1-H^2R^2}}{\sqrt{1-H^2R_0^{\ 2}}}\,
 \,\frac{\eps^2}{(x^2 + \epsilon^2)^{\frac{3}{2}}}\,\Big[1 + {\cal O} (x)\Big]
 = \frac{4 \pi R_0^{\ 2}\, \bar \rho}{3}  \,\frac{\eps^2}{(x^2 + \epsilon^2)^{\frac{3}{2}}}\,\Big[1 + {\cal O} (x)\Big]
 \label{epspperp}
\eea
where (\ref{mhr}), (\ref{R0div}), (\ref{rxdef}), (\ref{fexpand}), and (\ref{pperpreg}) have been used. 
The essential point is that the leading term in (\ref{epspperp}) cannot be discarded since although 
the one parameter sequence of regulated functions
\vspace{-1mm}
\be
\delta_{\eps}(x) \equiv \frac{1}{2}\, \frac{\eps^2}{(x^2 + \epsilon^2)^{\frac{3}{2}}}
\vspace{-3mm}\ee
satisfies
\vspace{-5mm}
\be
\lim_{\eps \rightarrow 0} \delta_{\eps}(x) = 0 \qquad {\rm for} \qquad x \neq 0\,,
\vspace{-1mm}
\ee
and thus gives no contribution at any finite $x$ in the limit, it also has the property that
\vspace{-2mm}
\be
\delta_{\eps}(0) = \frac{1}{2\eps}\qquad {\rm for\ any} \qquad \eps > 0
\vspace{-2mm}
\ee
\vspace{-2mm}\noindent
which is just that needed to give rise to a $\delta$-function distribution in the limit $\eps \rightarrow 0$.
Indeed since
\be
\int_{-\infty}^{\infty} dx \,\delta_{\eps}(x) = 1 \qquad \forall\, \eps
\ee 
the sequence of $\delta_{\eps}$ as $\eps \rightarrow 0^+$ defines precisely a $\delta$-distribution
\vspace{-1mm}
\be
\lim_{\eps \rightarrow 0} \delta_{\eps}(x) = \delta(x)\,.
\ee
In view of this result (\ref{epspperp}) becomes
\be
8\pi r^2  \left(\frac{f}{h}\right)^{\frac{1}{2}}\,\big(p_{\perp} - p\big) \rightarrow \frac{ 8\pi}{3}\,R_0^{\ 2}\, \bar\rho\ \delta(x) 
= \frac{ 8\pi}{3}\,R_0^{\ 3}\, \bar\rho\ \delta(r-R_0)
\label{delr0}
\ee
in the limit $\eps \rightarrow 0$, showing how a $\delta$-fn. contribution and breaking the isotropic perfect fluid 
condition localized on the singular surface at $r=R_0$ arises on the right side of (\ref{Mintegrand}), in agreement
with (\ref{lhslocr}).

The Komar energy associated with this surface tension is the integral of (\ref{delr0}) or $E_s$ given in (\ref{Esurf}).
Since $pf^{\frac{1}{2}}$ is regular at $r=R_0$, and the integration may be taken over an arbitrarily small region
surrounding $r=R_0$, the $\delta$-function contribution and the surface energy in (\ref{Esurf}) is attributable 
entirely to the transverse pressure term $p_{\perp}$, and therefore gives rise to a genuine surface tension.

One subtlety of this derivation is that the Taylor expansion of $D$ in (\ref{Dexpand}) breaks down and $c_1 \rightarrow \infty$
at the limiting point $R_0 = R= R_s$, because $D$ defined by (\ref{denom}) becomes a simple square root
and non-analytic in that limit. As a consequence, the metric factor $f(r) = \frac{1}{4} D^2$ vanishes {\it linearly} in $x$ 
at $r=R_s$ rather than quadratically as it does for all $R_0 < R_s$. Nevertheless since $c_1$ cancels from
the final combination (\ref{epspperp}), the $\delta$-function formula (\ref{delr0}) remains valid
in the limit $R \rightarrow R_s^+, R_0 \rightarrow R^-$. It is interesting to note that in this limit, even the
Principal Part prescription and regulator on the radial pressure in (\ref{preg}) become unnecessary, as
$p(r)$ contains no divergence, but only a step function discontinuity at $r=R_s$,\ {\it cf.} Fig. \ref{Fig:Pressureconst}.

We anticipate also that the formal regulator $\eps$ introduced in (\ref{preg}) and (\ref{fexpand}) will be replaced
by the dimensionless radial thickness $\Delta r/R_s$ of the shell with a small but finite proper width $\ell \sim \sqrt{R_s\Delta r}$
where $\Delta r$ is expected to be of order $L_{Pl}$ so that $\displaystyle \eps \sim \frac{L_{Pl}}{R_s} \sim \sqrt{\hbar}$
in a more complete quantum treatment.

\section{Junction Conditions and the Surface Tension of a Gravitational Condensate Star}
\label{Sec:Junction}

The matching of four-geometries across a general three-dimensional hypersurface interface with a 
$\delta$-function distribution of energy density and pressure concentrated on the interfacial boundary requires 
some care, particularly when the boundary becomes lightlike. In this Appendix we show how the matching determined
by interior Schwarzschild solution for $R \le \frac{9}{8} R_s$, continuity of the metric at $r=R_0$ and
the Komar mass-energy may be understood in the context of the general formalism of \cite{BarIsr}
and related to the present authors' previous work on gravitational condensate stars \cite{PNAS}.

In the case when the surface boundary is everywhere timelike, the normal to it $\bf n$ is spacelike and may be
normalized to unit length
\vspace{-4mm}
\be
{\bf n} \cdot {\bf n} = n^{\mu}g_{\mu\nu} n^{\nu} = +1
\label{spacen}
\vspace{-4mm}\ee
If the hypersurface of interest is at fixed $r$ in the Schwarzschild coordinates (\ref{sphsta}), this normal is
\be
n^{\mu} = \delta^{\mu}_{\ r} \sqrt{h(r)}\,.
\vspace{-3mm}\label{normal}
\ee
The mutually orthogonal basis vectors in the remaining three directions ${\bf \upsilon}_{(a)}$ have components
\vspace{-3mm}
\be
\upsilon_{(a)}^{\mu} = \delta^{\mu}_{\ a}\,,\qquad  a= t, \theta, \phi\,.
\label{upsdef}
\ee
The extrinsic curvature tensor is then defined by
\bea
{\bf K}_{ab} &\equiv &- n_{\mu}\, \upsilon_{(b)}^{\nu}\nabla_{\nu}\, \upsilon_{(a)}^{\mu}\nn
&=& - n^{\mu} \Gamma_{\mu ab} = - \sqrt {h}\, \Gamma_{r ab}
\label{Kdef}
\eea
with $\Gamma_{\mu\alpha\beta}$ the Christoffel symbol evaluated in the full four-metric, which is
required to be continuous (${\cal C}^0$) at $r=R_0$. On the other hand the extrinisic curvature is allowed 
to be discontinuous at the boundary, and its discontinuities
\be
[{\bf K}^a_{\ b}] \equiv {\bf K}^{a\, +}_{\ b} - {\bf K}^{a\, +}_{\ b} = 4 \pi G\, \Big( 2  S^a_{\ b} - \delta^a_{\ b} S^c_{\ c}\Big)
\label{Kdisc}
\ee
determine the surface stress-energy as a $\delta$-function distribution $^{(\Sigma )}T^a_{\ b} = S^a_{\ b}\,\delta(r-R_0)$
concentrated on the surface \cite{Isr}. Indices $a,b, c= \{t, \theta, \phi\}$ are raised and lowered with respect to the 
induced three-metric on $\Sigma$ which, because of  (\ref{upsdef}) is simply the four-metric of (\ref{sphsta}) restricted to $dr = 0$. 

Since from (\ref{Kdef})
\vspace{-3mm}
\be
{\bf K}^t_{\ t} = \frac{\sqrt{h}}{2f}\frac{df}{dr}\,,\qquad {\bf K}^{\theta}_{\ \theta} =  {\bf K}^{\phi}_{\ \phi} = \frac{\sqrt{h}}{r}
\label{Kcomp}
\ee
we have the surface stress-energy components\\
\vspace{-7mm}
\bes
\bea
-\bar S^t_{\ t} \equiv \eta &=& \frac{1}{4 \pi G} \left[\frac{\sqrt{h}\, }{r}\right]\\
\bar S^{\theta}_{\ \theta} = S^{\phi}_{\ \phi} \equiv \sigma &=& - \frac{1}{8 \pi G} \left\{ \left[ \frac{\sqrt{h\,}}{2f} \frac{df}{dr}\right] 
+  \left[\frac{\sqrt{h}}{r}\right]\right\}
\label{sigdef}\eea
\label{junc}\ees
with $\eta$ and $\sigma$ the surface energy density and surface tension in the rest frame of the surface
at $r=R_0$. From the matching of interior to exterior Schwarzschild solutions or Fig. \ref{Fig:hvarious} it is
clear that there is no discontinuity in $\sqrt{h(r)}$, which would give rise to a surface energy density $\eta= -\bar S^t_{\ t}$. 
Instead there is a rapid dropoff from constant  $\bar\rho$ to zero, which becomes a step function in the limit that the regulator of
Appendix \ref{Sec:DeltaDistrib}, $\eps \rightarrow 0$ but there is no $\delta$-function in the energy density $\rho$, and $\eta = 0$.

From the junction conditions (\ref{Kcomp})-(\ref{junc}) one sees that there is a breakdown of this formalism 
when the surface becomes lightlike due to $f \rightarrow 0$. Although the discontinuity in $\sqrt{h(r)}$ may be assumed to be
small or identically zero so that the surface energy density $\eta$ vanishes, the surface tension in the rest
frame of the surface $\sigma$ diverges. This is clearly a simple kinematic effect of the redshift factor $f^{\frac{1}{2}}$ 
going to zero on a null hypersurface, where an infinite boost would be required and no rest frame exists.
In the interior Schwarzschild solution this occurs at $r=R_0$ where $f=0$ for any value of $0 \le R_0 \le R_s$,
and in particular persists in the limit $R_0 \rightarrow R_s$, when $h(R_0) \rightarrow 0$ as well, and
the surface coincides with the Schwarzschild sphere at $r=R_s$.

Because of the apparent difficulty in joining a de Sitter interior to a Schwarzschild exterior at their common null boundaries 
at $H = 1/R_s$, a maximally stiff surface layer of $p= \rho$ fluid was assumed to be interposed between the de Sitter interior 
and Schwarzschild exterior solutions in \cite{PNAS}. In that case there are two separate timelike boundaries at $r=r_1$ and $r=r_2$ 
with $r_1< r_2$ both very close to the horizon, where the Isreal junction conditions (\ref{Kcomp})-(\ref{junc}) can be applied
unmodified. Approached from the de Sitter interior side the extrinsic curvature at $r= r_1$ in \cite{PNAS} is
\be
{\bf K}_t^{\ t}(r_1^-) =  \frac {\sqrt h}{2} \frac {d }{dr} \ln (1- H^2 r^2)\big\vert_{r= r_1^-} = - \frac{H^2r_1}{\sqrt{1-H^2r_1^2}} \rightarrow 
- \frac{1}{\sqrt{1-H^2r_1^2}}\frac{1}{R_s}
\ee
as $Hr_1 \rightarrow 1$. Since $h(r_1) =1-H^2r_1^2$ is of order $\eps$, with $\Delta r = r_2 - r_1 = \eps R_s \sim L_{Pl}$
this extrinsic curvature component and its discontinuity at $r_1$ is of order $\eps^{-\frac{1}{2}}$. This accounts for
the $\eps^{-\frac{1}{2}}$ dependence of the surface tensions $\sigma_1$ and $\sigma_2$ at each of the two surface 
layers in \cite{PNAS}. However the surface gravity $\kappa(r)$ differs from ${\bf K}_t^{\ t}(r)$ by a factor of 
$\sqrt{f(r)} = \frac{1}{2} \sqrt{h(r)}$ in the Schwarzschild interior solution (\ref{fsoln}), so that
\vspace{-3mm}
\be
\kappa(r_1^-)  = \sqrt{f(r_1^-)}\ {\bf K}_t^{\ t}(r_1^-)  = -\frac{1}{2R_s} = \kappa_-
\ee
remains finite, in agreement with (\ref{kappm}) at $R_0 = R = R_s$. The factor of $\sqrt{f(r)}$ is the Tolman redshift factor 
transforming the local extrinsic curvature and surface tension in the surface rest frame to the time global Killing time
$t$ of the static spacetime.

Likewise at the outer Schwarzschild boundary we have 
\be
{\bf K}_t^{\ t}(r_2^+) =  \frac {\sqrt h}{2} \frac {d}{dr} \ln \left(1 - \frac{R_s}{r}\right)\bigg\vert_{r= r_2^+} =
\frac{1}{2\sqrt{1- R_s/r_2^+}}\frac{1}{R_s}
\ee
which is again of order $\eps^{-\frac{1}{2}}$. Multiplying by the exterior Schwarzschild solution redshift factor 
we obtain the surface gravity\\
\vspace{-1.1cm}
\be
\kappa(r_2^+)  = \sqrt{f(r_2^+)}\ {\bf K}_t^{\ t}(r_2^+)  = +\frac{1}{2R_s} =  \kappa_+
\ee
which is again in agreement with (\ref{kappm}) at $R_0 = R = R_s$. 

Since these surface gravities including the redshift factor $\sqrt{f(r)}$ remain finite in the limit $\eps \rightarrow 0$ 
as the boundary surfaces approach each other at the common de Sitter-Schwarzschild horizon, 
one can take the limits $r_1^-  \rightarrow R_s, r_2^+ \rightarrow R_s$, dispensing with the interposition of a
surface layer of $p= \rho$ fluid entirely and define the surface tension by (\ref{deftau}), thus obtaining (\ref{taus}).
In \cite{PNAS} $\eta \propto \eps^{\frac{1}{2}} \rightarrow 0$ in agreement with the results of this work.
In this limit therefore one obtains from the interior Schwarzshild solution a {\it universal} result for
the (redshifted) surface tension of a gravitational condensate star, free of any model dependent assumptions 
of the surface layer eq. of state or other properties or parameters. 

This result may be connected with the general formalism of matching across null surface interfaces
described in \cite{BarIsr} as follows. These authors define a modified or `oblique' extrinsic curvature
\be
{\cal K}_{ab} =  -N_{\mu}\, \upsilon_{(b)}^{\nu}\nabla_{\nu}\, \upsilon_{(a)}^{\mu}
\ee
where $\bf N$ is an oblique or transversal null vector satisfying
\be
{\bf N}\cdot {\bf N} = 0\,,\qquad {\bf n} \cdot {\bf N} = -1
\label{Ncond}
\vspace{-3mm}
\ee
and the limit 
\vspace{-6mm}
\be
{\bf n} \cdot {\bf n} = \eps \rightarrow 0\,.
\label{ncond}
\ee
is taken. Thus the normal vector $\bf n$ is allowed to approach a null vector as well, in contrast to the uniform
spacelike condition (\ref{spacen}). By a judicious choice of this normalization and the transversal
vector $\bf N$ and application of the junction conditions to ${\cal K}_{ab}$, it is possible to obtain
well-defined finite results for matching on null surfaces \cite{BarIsr}. Indeed, if one one chooses (for example)
\bes
\begin{align}
\vspace{-4mm}
n^{\mu} &= \sqrt{hf\,} \, \delta^{\mu}_{\ r}\,,\qquad {\bf n}\cdot{\bf n} = f \sim \eps \rightarrow 0 \\
N^{\mu} &= \frac{1}{f}\, \delta^{\mu}_{ t} - \sqrt{\frac{h}{f}}\ \delta^{\mu}_{\ r} 
\end{align}
\label{nNchoice}\ees
satisfying (\ref{Ncond}), one finds
\vspace{-5mm}
\be
{\cal K}_{ab} = - N^{\mu}\, \Gamma_{\mu ab} = - \sqrt{\frac{h}{f}} \ \Gamma_{r ab} = f^{- \frac{1}{2}}\, {\bf K}_{ab}\,.
\label{oblique}
\ee
Note that together with (\ref{Kcomp}) this implies that any discontinuity of $h(r)$ at the Killing horizon
$r=R_0$, where $f(r) = 0$ would lead to a linear divergence in ${\cal K}^{\theta}_{\ \theta} = {\cal K}^{\phi}_{\ \phi}$.
The constant density interior Schwarzschild solution avoids this potential problem for any $R_0 \le R_s$ 
by the continuity of the metric function $h(r)$ and $\eta=0$ at $r=R_0$.

The modified or oblique extrinsic curvature (\ref{oblique}) has the $tt$ component
\be
{\cal K}_{tt} = -\frac{1}{2} \sqrt{\frac{h}{f}}\, \frac{df}{dr} = - \kappa
\ee
which is discontinuous at $r=R_0$, leading to the junction condition
\be
{\cal S}^A_{\ B} = \frac{[\kappa]}{8 \pi G} \, \delta^A_{\ B} = \tau_s \, \delta^A_{\ B}\,,\qquad A, B = {\theta, \phi}
\label{SAB}
\ee
assuming no discontinuity in $\sqrt{h(r)}$.  In the conventions of \cite{BarIsr} one finds then that
\be
^{(\Sigma )}T^A_{\ B} = \sqrt{\frac{h}{f}}\ {\cal S}^A_{\ B}\,\delta(r-R_0)\,,\qquad  ^{(\Sigma )}T^A_{\ B}\, \sqrt{\frac{f}{h}} 
= \tau_s \,\delta^A_{\ B}\,\delta(r-R_0)
\label{Tdeltrans}
\ee
is the distributional stress tensor density on the null surface with only transverse $T^{\theta}_{\ \theta} = T^{\phi}_{\ \phi}$
components. The last form of (\ref{Tdeltrans}) is precisely the term in the integrand in the Komar mass-energy
(\ref{locr}) or (\ref{Mintegrand}) giving a $\delta$-function which is integrable with respect to $r$ with the correct
measure factor, and (\ref{SAB}) corresponds exactly to the redshifted surface tension of the null surface (\ref{deftau}) 
and (\ref{taus}), obtained by application of the Komar integral formula to the interior Schwarzschild solution. 

In addition to the redshift factor $f^{\frac{1}{2}}$ defining the surface gravity and surface tension with respect to the Killing 
time $t$ rather than the proper time of the shell, note that $\tau_s$ and $\sigma$ of (\ref{sigdef}) have {\it opposite} signs.
Whereas the usual surface tension of a bubble membrane supplies an inward force (corresponding to a negative 
$p_{\perp}$) resisting expansion of the enclosed volume with positive pressure, and in the usual case the pressure
gradient $\displaystyle \frac{dp}{dr}$ is negative at the boundary, the positive $\tau_s$ and transverse pressure
$p_{\perp} > 0$ of the gravitational condensate surface produces an {\it outward} force, tending to {\it expand}
the surface area, and the pressure gradient $\displaystyle \frac{dp}{dr}$ is {\it positive} at the boundary.
Thus we have finally
\bes
\bea
\vspace{-2cm}
&&\eta = 0\\
&&\tau_s = - f^{\frac{1}{2}} \sigma
\eea
\ees
for the relationship of the present work to ref. \cite{PNAS}, with $\sigma$ that obtained by the {\it total} discontinuity
across the full surface layer from Schwarzschild exterior ($r_2^+$) to modified de Sitter interior ($r_1^-$).  Using
(\ref{kapxi}) one may easily check that the same discontinuity (\ref{SAB}) and $\delta$-function
distributional transverse stress tensor (\ref{Tdeltrans}) is obtained in the Rindler-like coordinates (\ref{txi}).

We remark finally that these results, although satisfactory, depend upon the choices (\ref{nNchoice}) which are
not unique given the conditions (\ref{Ncond}) and (\ref{ncond}). The crucial condition (\ref{nNchoice})
satisfies is 
\be
{\bf n} \cdot {\bf n} = f= -K_{\mu}K^{\mu}
\ee
which is the (negative) norm of the timelike Killing vector of stationarity on both sides of the singular surface,
normalized by referral to the asymptotic time at infinity. It is this normalized Killing field that plays a privileged 
role in the Komar mass-energy and surface gravity
\be
\kappa = - N_{\mu} K^{\nu} \,\nabla_{\nu} K^{\mu} = - N^{\mu} \, \Gamma_{\mu  tt} = \frac{1}{2} \sqrt{\frac{h}{f}}\, \frac{df}{dr}\,.
\ee
Rescalings ${\bf n} \rightarrow \lambda {\bf n}, {\bf N}  \rightarrow  \lambda^{-1}{\bf N}, \eps \rightarrow \lambda^2 \eps$, 
with $|\lambda| \neq 1$, which would be consistent with (\ref{Ncond})-(\ref{ncond}) in the general formalism of \cite{BarIsr} are 
thereby disallowed by the physical requirement of matching the preferred asymptotic stationary Killing time $t$ on both sides 
of the surface.

\end{document}